\documentclass{article}
\usepackage[a4paper, total={6in, 8in}]{geometry}
\usepackage[indent,skip=4pt]{parskip}

\usepackage{amsmath,amssymb,amsfonts,amsthm,mathrsfs,mathtools,cases}
\usepackage{dsfont}
\usepackage{authblk}
\usepackage[colorlinks=true,linkcolor=blue, citecolor=blue, bookmarks]{hyperref}
\usepackage{cleveref}
\usepackage{microtype}
\usepackage{comment}
\usepackage{graphicx} 
\usepackage{subfig}
\usepackage{wrapfig}
\usepackage[font=small,labelfont=bf]{caption}
\usepackage[usenames,dvipsnames]{xcolor}  
\usepackage{booktabs}
\usepackage{braket}
\usepackage{nicematrix}
\usepackage{tikz}
\usepackage{pgfplots}
\usepackage{cite}
\pgfplotsset{compat=1.18}

\usetikzlibrary{matrix}
\usetikzlibrary{decorations.markings}

\newcommand{\II}{\mathbb I}
\newcommand{\RR}{\mathbb R}

\newcommand{\ZZ}{\mathbb Z}

\newcommand{\hilb}{\mathcal H}

\newcommand{\boldg}{\mathbf g}
\newcommand{\boldh}{\mathbf h}

\newcommand{\bx}{\mathbf x}
\newcommand{\Zng}{\mathcal Z_n(\boldg)}

\newcommand{\hess}{\mathsf H}
\newcommand{\hessTn}{\mathsf H_{T_n}}

\newcommand{\hessbetan}{\mathsf H_{\beta_n}}

\newcommand{\id}{\mathds 1}
\DeclareMathOperator{\vol}{vol}

\DeclareMathOperator{\res}{res}
\DeclareMathOperator{\Log}{Log}
\DeclareMathOperator{\tr}{tr}
\DeclareMathOperator{\arctanh}{arctanh}
\renewcommand*\d{\mathop{}\!\mathrm{d}} 

\newcommand{\llambda}{\boldsymbol{\lambda}}
\newcommand{\aalpha}{\boldsymbol{\alpha}}
\newcommand{\lr}[1]{\left( {#1} \right)}
\newcommand{\slr}[1]{\left[{#1} \right]}
\newcommand{\clr}[1]{\left\{{#1} \right\}}

\crefname{figure}{Fig.}{Figs.}
\crefname{equation}{Eq.}{Eqs.}
\crefname{section}{Sec.}{Secs.}
\crefname{appendix}{Appendix}{Appendices}
  
\colorlet{darkerblue}{MidnightBlue!70!black}
\colorlet{lightblue}{blue!70!white}

\hypersetup{
    colorlinks,
    linkcolor={black},
    citecolor={black},
    urlcolor={darkerblue},
}

\title{\textbf{Entanglement asymmetry in CFT and its relation to non-topological defects}}
\author[1]{Michele Fossati}
\author[1]{Filiberto Ares}
\author[2,3]{Jerome Dubail}
\author[1,4]{Pasquale Calabrese}

\affil[1]{\textit{SISSA and INFN, Via Bonomea 265, 34136 Trieste, Italy}}
\affil[2]{\textit{CESQ and ISIS, CNRS \& University of Strasbourg, UMR 7006, 67000 Strasbourg, France}}
\affil[3]{\textit{LPCT, CNRS \& Universit\'e de Lorraine, UMR 7019, 54000 Nancy, France}}
\affil[4]{\textit{ICTP, Strada Costiera 11, 34151 Trieste, Italy}}

\begin{document}
\maketitle
\begin{abstract}
The entanglement asymmetry is an information based observable that quantifies the degree of symmetry breaking in a region of an extended quantum system. We investigate this measure in the ground state of one dimensional critical systems described by a CFT. Employing the correspondence between global symmetries and defects, the analysis of the entanglement asymmetry can be formulated in terms of partition functions on Riemann surfaces with multiple non-topological defect lines inserted at their branch cuts. For large subsystems, these partition functions are determined by the scaling dimension of the defects.
This leads to our first main observation: at criticality, the entanglement asymmetry acquires a subleading contribution scaling as $\log \ell / \ell$ for large subsystem length $\ell$. Then, as an illustrative example, we consider the XY spin chain, which has a critical line described by the massless Majorana fermion theory and explicitly breaks the  $U(1)$ symmetry associated with rotations about the $z$-axis. In this situation the corresponding defect is marginal. Leveraging  conformal invariance, we relate the scaling dimension of these defects to the ground state energy of the massless Majorana fermion on a circle with equally-spaced point defects. We exploit this mapping to derive our second main result: the exact expression for the scaling dimension associated with $n$ of defects of arbitrary strengths. Our result generalizes a known formula for the $n=1$ case derived in several previous works. We then use this exact scaling dimension to derive our third main result: the exact  prefactor of the $\log \ell/\ell$ term in the asymmetry of the critical XY chain.
\end{abstract}

\newpage
\tableofcontents
\newpage

\section{Introduction}
Symmetries play a pivotal role in the foundations of modern physics. 
Their presence implies conservation laws that have deep 
consequences in the behavior of physical systems and facilitate 
enormously the resolution of many problems, which  would 
otherwise remain open. As crucial as the existence of symmetries is their 
breaking, both explicit and spontaneous. 
Such breaking is responsible for a plethora of very important phenomena across different 
branches of physics. A relevant aspect that has received little 
attention so far is the quantification of how much a global symmetry 
is broken. Local order parameters have been usually employed to 
discern whether or not a quantum state respects a symmetry. 
However, they present  the disadvantage that, while a non-zero value 
manifests that the symmetry is broken, the converse is not always 
true. Furthermore, in extended quantum systems, the question of 
measuring symmetry breaking is intrinsically tied to consider a 
specific subsystem. In fact, there may exist long-range correlations 
between the parts of the system that do not respect the symmetry and 
are not taken into account by any local order parameter. 

In this context, an appealing idea is quantifying symmetry breaking 
by leveraging tools from the theory of entanglement, as they capture 
non-local correlations. A quantity based on the entanglement entropy 
and dubbed entanglement asymmetry has been recently introduced as a 
measure of how much a symmetry is broken in a subsystem. The 
entanglement asymmetry has proven to be a powerful instrument 
to identify novel physical phenomena. 
It has been applied to 
investigate the dynamical restoration of a $U(1)$ symmetry from an initial 
state that breaks it after a quench to a Hamiltonian that respects 
the symmetry~\cite{amc-23}. Surprisingly, the entanglement asymmetry shows that 
the restoration of the symmetry may occur earlier for those states 
that initially break it more, a quantum version of the yet 
unexplained Mpemba effect (the more a system is out of equilibrium, 
the faster it relaxes). This quantum Mpemba effect has been observed 
experimentally by measuring the entanglement asymmetry in an ion 
trap~\cite{joshi-24} and the microscopic mechanism and the conditions 
under which it occurs are now well understood for free and interacting integrable 
systems~\cite{rkacmb-23, Murciano:asymmetryXY, bkccr-23}, although 
they remain elusive for non-integrable ones. In addition, the 
entanglement asymmetry has been applied to examine the dynamical 
restoration of a spontaneously broken $\mathbb{Z}_2$ symmetry~\cite{fac-23} and the 
relaxation to a non-Abelian Generalized Gibbs ensemble in the exotic 
case that the symmetry is not restored~\cite{amvc-23}. It has 
been also generalized to study the quench dynamics of 
kinks~\cite{kkhpk-23}. Beyond non-equilibrium physics, the 
entanglement asymmetry has been employed to understand the 
implications of quantum unitarity for broken symmetries during black hole evaporation~\cite{ampc-23}.

A significant point in the characterization of the entanglement 
asymmetry is its asymptotic behavior with the size of the subsystem 
considered. As this observable is based on the entanglement entropy, 
one may wonder whether it inherits some of its properties. 
For example, 
the entanglement entropy follows an area law in the ground
state of one dimensional systems with mass gap.
In contrast it grows 
logarithmically with the subsystem size when the mass gap vanishes; 
this logarithmic growth is  proportional to the central charge of the conformal field 
theory (CFT) that describes the low energy physics of the critical point~\cite{hlw-94, cc-04,cc-09}. 
Conversely, the entanglement asymmetry exhibits a fundamentally distinct behavior. 
It has been shown in Ref.~\cite{CapizziVitaleMPS:2023} that, for matrix product 
states, the entanglement asymmetry for a generic compact Lie 
group grows at leading order logarithmically with the
subsystem size, with a coefficient proportional to the dimension of 
the Lie group, while, for finite discrete groups, the 
entanglement asymmetry satisfies an area law, saturating to a value 
fixed by the cardinality of the group. Similar results
have been obtained in the ground state of the XY spin chain when 
studying the particle number $U(1)$ symmetry that this model 
explicitly breaks~\cite{Murciano:asymmetryXY} and the spin-flip $\mathbb{Z}_2$ symmetry, spontaneously broken in the ferromagnetic phase~\cite{fac-23, CapizziMazzoniIsing:2023}.

In this paper, we examine the implications of quantum criticality for 
the entanglement asymmetry, which remain barely unexplored, using CFT 
methods. 
Only Ref.~\cite{chen-chen-23} reports calculations for the entanglement 
asymmetry in certain particular excited states of the massless compact boson. 
To this end, we develop a general scheme to compute the entanglement 
asymmetry in (1+1)-dimensional quantum field theories in terms of the 
charged moments of the subsystem's reduced density matrix. Employing 
the path integral formulation, the  charged moments can be identified 
with the partition functions of the theory on Riemann surfaces with 
defect lines inserted along its branch cuts. These defect lines are 
associated with the elements  of the symmetry group under 
analysis~\cite{ffrs-04,gksw-15}. A symmetry is considered broken when the associated defects are not topological, and any continuous deformation of these defects leads to a change in the partition function. Therefore, within this framework, the 
entanglement asymmetry can be naturally interpreted as a measure of 
how much the defects are not topological. We apply this approach to 
determine the entanglement asymmetry in the ground state of the XY 
spin chain at the Ising critical line for the $U(1)$ group of spin rotations around the transverse direction. After fermionizing it through a Jordan-Wigner 
transformation, the scaling limit of this model is described by the 
massless Majorana fermion theory and the defect lines corresponding to this group are marginal. We then exploit conformal invariance to map the Riemann surfaces to a single cylinder with defect lines parallel to its axis. In this setup, the calculation of the partition functions for large subsystems boils down to 
computing the ground state energy of the massless Majorana 
fermion on a circle with equally-spaced marginal point defects. The 
spectrum of this theory has been studied on the lattice in Refs.~\cite{hp-88, hps-89}. 
Here we revisit this problem and diagonalize systematically its 
Hamiltonian for an arbitrary number of equi-spaced point defects of
different strengths. The study of defects in the massless Majorana 
fermion and Ising CFTs has a long story, see e.g.~\cite{turban-85, ipt-93, oa-96, oa-97, pz-00, ffrs-04, qrw-07, ffrs-07, ffn-09, bbr-13, amf-16, afm-20, mslma-23}. Partition functions on Riemann 
surfaces with (topological and non topological) defect lines also arise in the analysis of the entanglement across inhomogeneities, interfaces, or junctions and after measurements~\cite{ss-08, ep-10, cmv-12, ep-12, bb-15, gm-17,  mt-21, cmc-22, cmc-23, cr-23, ymj-23,wsag-23, ljx-23, afo-23, rcgf-22}; in particular, those with 
topological defect lines appear in the symmetry resolution of entanglement measures~\cite{goldstein, xavier, cgs-18, mdgc-20, cdm-21, crc-20, chen-21, cc-21, brc-19, bc-21, eim-21, mbc-21, mt-23, acdgm-22, chen-22, ghasemi-23, dgmnsz-23,  gy-23, fmc-23, northe-23, kmop-23, bamc-23, fac-23-2}, which has recently been investigated in
profusion.

The paper is organized as follows. In Sec.~\ref{sec:topological-defects-asymmetry}, we review the 
relation between symmetries and defects in (1+1)-quantum field 
theories, we introduce the entanglement asymmetry, and we show
how to compute it from the partition function on a Riemann surface 
with defect lines. We also derive the asymptotic behavior of the 
entanglement asymmetry for a generic compact Lie group 
in the ground state of a one dimensional critical system. In the rest 
of the sections, we focus on the critical XY spin chain and the associated CFT, the massless Majorana fermion theory. In 
\cref{sec:setup}, we introduce these systems and we review the known previous 
results for the entanglement asymmetry. In Sec.~\ref{sec:CFT_calculation_Delta}, we calculate the 
partition function of the Majorana CFT on the Riemann 
surfaces that enter in the calculation of the entanglement asymmetry. 
In particular, by conformal invariance, these partition functions are 
given by the ground state energy of a massless Majorana fermion with 
evenly-spaced point defects. We carefully diagonalize its Hamiltonian 
for an arbitrary number of defects with different strengths. In \cref{sec:asymm_XY}, we apply these results to obtain the entanglement asymmetry
of the critical XY spin chain, checking them against exact numerical 
computations on the lattice. Finally, in Sec.~\ref{sec:conclusions}, we draw our 
conclusions and consider future prospects. We also include several 
appendices where we discuss with more detail some technical points of 
the main text.

\section{Symmetries, topological defects, and entanglement asymmetry}\label{sec:topological-defects-asymmetry}
In this section, we briefly review the identification between 
symmetries and topological defects. Then we introduce the R\'enyi 
entanglement asymmetry as a quantifier of symmetry breaking and we 
interpret it in terms of defects. With simple scaling arguments, we 
derive some general results for the asymptotic behavior of the 
R\'enyi entanglement asymmetry in the ground state of a critical 
one dimensional quantum system in the thermodynamic limit.

\subsection{Symmetries and topological defects}\label{sec:symm_top_def}

Global symmetries in spatially extended quantum systems are realized through extended operators that form a unitary representation of the symmetry group. In fact, if we consider a generic (1+1)-dimensional quantum field theory whose spacetime is a flat surface $\mathcal{M}$, then the action of an 
element $g$ of the group $G$ (either discrete or continuous) is implemented in its Hilbert space $\mathcal{H}$ by a unitary operator $U_{\Sigma_t, g}$ that has support on a spatial line $\Sigma_t\subset \mathcal{M}$ at a fixed time $t$. A familiar instance is the case of a $U(1)$ symmetry. The Noether theorem  ensures the existence of a conserved current $j^\mu$. Therefore, the associated charge at $\Sigma_t$ is $Q_{\Sigma_t}= \int_{\Sigma_t} \d x \,j^0(x)$ and the group is represented by the operators $U_{\Sigma_t, \alpha} = \exp [i \alpha Q_{\Sigma_t}]$, with $\alpha\in[0, 2\pi)$. 

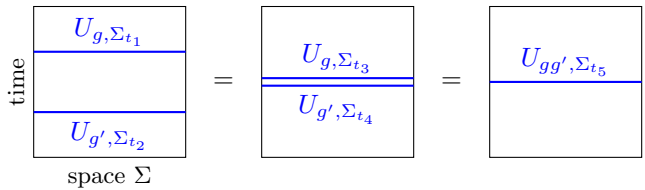
\begin{figure}[t]
\centering
    \begin{tikzpicture}
  \draw[] (0,0) -- (2,0) -- (2,2) -- (0,2) -- cycle;
  
  \draw[blue, thick] (0,0.6) -- (2,0.6);
  \draw[blue, thick] (0,1.4) -- (2,1.4);
  
  \node[below, blue] at (1,0.6) {$U_{g',\Sigma_{t_2}}$};
  \node[above, blue] at (1,1.35) {$U_{g,\Sigma_{t_1}}$};
  \node[below] at (1,0) {\small space $\Sigma$};
  \node[above,rotate=90] at (0,1) {\small time};

    \node at (2.5,1) {$=$};
    \node at (5.5,1) {$=$};
    
    \begin{scope}[xshift=3cm]
  \draw[] (0,0) -- (2,0) -- (2,2) -- (0,2) -- cycle;
  
  \draw[blue, thick] (0,0.95) -- (2,0.95);
  \draw[blue, thick] (0,1.05) -- (2,1.05);
  
  \node[below, blue] at (1,0.95) {$U_{g',\Sigma_{t_4}}$};
  \node[above, blue] at (1,1.0) {$U_{g,\Sigma_{t_3}}$};
  
\end{scope}

\begin{scope}[xshift=6cm]
  \draw[] (0,0) -- (2,0) -- (2,2) -- (0,2) -- cycle;

  \draw[blue, thick] (0,1) -- (2,1);
  
  \node[above,blue] at (1,0.95) {$U_{gg',\Sigma_{t_5}}$};
\end{scope}
\end{tikzpicture}
\caption{Each element $g$ of a group $G$ acts on the Hilbert space of an extended quantum system as a unitary operator $U_{\Sigma_t, g}$ defined along a line $\Sigma_t$ at a fixed time $t$. If $G$ is a symmetry of the theory, then any continuous  transformation of $\Sigma_t$, as the ones performed in the figure, leaves invariant the partition function with insertions of these operators. We indicate this by the symbol $=$ between the three diagrams. When two operators $U_{\Sigma_t, g}$ and $U_{\Sigma_t, g'}$ overlap, as in the right diagram, they can be fused according to the composition rule $U_{\Sigma_t, g}U_{\Sigma, g'}=U_{\Sigma_t, gg'}$.}
\label{fig:fusion-lines}
\end{figure}

The extended operators $U_{\Sigma_t, g}$ representing symmetries possess the crucial property of being \emph{topological}. This means that continuous deformations of $\Sigma_t$ do not affect any expectation value that contains the insertion of an operator $U_{\Sigma_t, g}$. For example, since a symmetry operator commutes with the Hamiltonian of the theory, it will not evolve in the Heisenberg picture and then $U_{\Sigma_t, g} = U_{\Sigma_{t'}, g}$, as depicted in the first equality of \cref{fig:fusion-lines}. When the support of two extended operators $U_{\Sigma_t, g}$, $U_{\Sigma_t, g'}$ coincides, the operators fuse according to the standard composition rule $U_{\Sigma_t, g}U_{\Sigma_t, g'}=U_{\Sigma_{t}, gg'}$, as we illustrate in the second equality of \cref{fig:fusion-lines}. 

The transformation of a field $\phi$ of the theory under the group $G$ is described by a matrix $R_{g}$ such that
\begin{equation}
 U_{\Sigma_t, g}^\dagger\phi(x) U_{\Sigma_t, g}=R_g\phi(x), \quad x\in\Sigma_t.
\end{equation}
Therefore, within the path integral formalism, the insertion of an operator $U_{\Sigma_t, g}$ in an expectation value is equivalent to performing a cut along the line $\Sigma_t$ and imposing for the fields the following gluing conditions 
\begin{equation}\label{eq:defect-path-integral}
\phi(x^+)=R_g \phi(x^-), \quad x\in \Sigma_t,
\end{equation}
where $\phi(x^\pm)$ denote the field $\phi(x)$ at each side of the cut as we indicate in Fig.~\ref{fig:defect-path-integral}. The composition property $U_{\Sigma_t, g}U_{\Sigma_t, g'}=U_{\Sigma_t, gg'}$ can be then 
understood as the fusion of two cuts with gluing conditions $R_{g}$ and $R_{g'}$ into
a cut with gluing condition $R_{g}R_{g'}=R_{gg'}$.
In Euclidean spacetime, $U_{\Sigma, g}$ is not needed to be defined
along a line $\Sigma_t$ orthogonal to the time direction, but it can have support on any curve $\Sigma$ 
on the surface $\mathcal{M}$. Due to the previous considerations,
the extended operators $U_{\Sigma, g}$ are commonly referred to as {\it defects}, and when they enforce symmetries,
they are {\it topological defects}~\cite{gksw-15}. A more detailed introduction to the role of topological operators in quantum systems can be found in, e.g., the recent review~\cite{bhardwaj2023lectures}.

\begin{figure}[t]
    \begin{center}
    \begin{tikzpicture}
    \draw[blue, thick] (0,1) -- (2,1); 
    \node[left, blue] at (0,1) {$U_{\Sigma, g}$};
    \fill (1,0.9) circle (1pt);
    \node[below left] at (2,1) {$R_g\phi(x^-)$};
    \node at (2.5,1) {$=$};

    \begin{scope}[xshift=3.2cm]
        \draw[blue, thick] (0,1) -- (2,1);
        \node[right, blue] at (2,1) {$U_{\Sigma, g}$};
        \fill (1,1.1) circle (1pt);
        \node[above left] at (1.6,1) {$\phi(x^+)$};
    \end{scope}
\end{tikzpicture}
\end{center}
\caption{Graphical representation of Eq.~\eqref{eq:defect-path-integral}. The insertion of an extended operator $U_{\Sigma, g}$ associated with the element $g$ of a group $G$ and with support on the line $\Sigma$ corresponds, in the path integral approach, to a defect line along $\Sigma$ with the gluing condition~\eqref{eq:defect-path-integral} for the field $\phi(x)$ at each side of the defect.}
\label{fig:defect-path-integral}
\end{figure}

The question of whether a system is symmetric under a certain group 
can thus be reformulated as asking whether the defects associated to 
the symmetry are topological. In this paper, we are interested
in quantifying the extent to which a symmetry is broken or, in other 
words, measuring how much the corresponding defects are not 
topological. This can be done with the entanglement asymmetry, which 
we now introduce.

\subsection{Entanglement asymmetry}\label{sec:asymm_def}
\subsubsection{Definition}
Let us take an extended quantum system in a state described by the density matrix $\rho$. We consider a spatial bipartition $\Sigma=A\cup \bar{A}$ in which $A$ consists of a single connected region such that the total Hilbert space $\hilb$ factorizes into $\mathcal{H}=\mathcal{H}_A\otimes \mathcal{H}_{\bar A}$. We assume that the extended operators that represent the group $G$ decompose accordingly as $U_{\Sigma , g}=U_{A, g}\otimes U_{\bar{A}, g}$. The state of subsystem $A$ is given by the reduced density matrix $\rho_A= \tr_{\bar{A}}\rho$, obtained by tracing out the degrees of freedom in the region $\bar{A}$. Under an element of the group $G$, it transforms as $\rho_A\mapsto U_{A, g}\rho_A U_{A, g}^\dagger$. Therefore, the state $\rho_A$ is symmetric if $[\rho_A, U_{A, g}]=0$ for all $g\in G$.

To define the entanglement asymmetry, we introduce the \emph{symmetrization} of $\rho_A$ as the average over $G$ of the transformed density matrix $U_{A, g}\rho_AU_{A, g}^\dagger$; that is,
\begin{equation}\label{eq:def-symmetrized-Lie}
    \rho_{A,G} \coloneqq \frac{1}{\vol G}\int_G \d g  \, U_{A,g} \rho_A U_{A,g}^\dagger,
\end{equation}
if $G$ is a compact Lie group, where $\d g$ is its Haar measure and $\vol G$ its volume. An analogous formula can be written up for a finite discrete group $G$ of cardinality $|G|$  replacing the Haar integral by a sum over its elements. To lighten the discussion, we focus on compact Lie groups and we refer the reader to Refs.~\cite{fac-23, CapizziMazzoniIsing:2023, CapizziVitaleMPS:2023} where the entanglement asymmetry has been examined for discrete groups.  The density matrix $\rho_{A, G}$ is by construction
symmetric under $G$ and has trace one. Note that $\rho_A$ is symmetric if and only if $\rho_{A}=\rho_{A, G}$. Then the entanglement asymmetry is the relative entropy between $\rho_A$ an $\rho_{A, G}$~\cite{amc-23},
\begin{equation}
 \Delta S_A:=S(\rho_A||\rho_{A, G})=\tr\left[\rho_A(\log\rho_{A}-\log\rho_{A, G})\right].
\end{equation}
Given the form of $\rho_{A, G}$, and applying the cyclic property of the trace, $\Delta S_A$ can be rewritten as
\begin{equation}\label{eq:vn_ent_asymm}
 \Delta S_A=S(\rho_{A, G})-S(\rho_A),
\end{equation}
where $S(\rho)$ is the von Neumann entropy of $\rho$, $S(\rho) =-\tr(\rho \log \rho)$. The entanglement asymmetry satisfies two essential properties as a measure of symmetry breaking in the subsystem $A$: it is non-negative, $\Delta S_A\geq 0$, and it vanishes if and only if $A$ is in a symmetric state, i.e.  $\rho_A=\rho_{A, G}$ \cite{Nielsen-Chuang, mhms-22}.

In general, the direct calculation of the entanglement asymmetry is complicated due to the presence of the logarithm in the von Neumann entropy. Alternatively, a much simpler indirect way of computing it is via the replica trick~\cite{hlw-94, cc-04,cc-09}. By replacing in Eq.~\eqref{eq:vn_ent_asymm} the von Neumann entropy
by the R\'enyi entropy, $S^{(n)}(\rho)=\frac{1}{1-n}\log\tr\rho^n$, we introduce the R\'enyi entanglement asymmetry
\begin{equation} \label{eq:asymmetry-difference-reny}
    \Delta S_A^{(n)} = S^{(n)}(\rho_{A,G}) - S^{(n)}(\rho_A).
\end{equation}
Observe that the entanglement asymmetry~\eqref{eq:vn_ent_asymm} is recovered in the limit 
$\displaystyle \lim_{n\to 1}\Delta S_A^{(n)}=\Delta S_A$. The advantage of the R\'enyi entanglement asymmetry is that, for integer $n$, it can be expressed in terms of charged partition functions. If we plug the definition of $\rho_{A, G}$ in \cref{eq:asymmetry-difference-reny}, we obtain 
\begin{equation}\label{eq:renyi_ent_asymm_charged_mom_no_dirac_delta}
    \Delta S_A^{(n)} = \frac{1}{1-n} \log \slr{ \frac{1}{(\vol G)^n} \int_{G^n} \d \boldg \frac{\tr(U_{A,g_1} \rho_A U_{A,g_1^{-1}} \dots U_{A,g_n} \rho_A U_{A,g_n^{-1}})}{\tr(\rho_A^n)}}, 
\end{equation}
where $G^n= G \times \stackrel{n}{\cdots} \times G$ and $\boldg$ stands for the $n$-tuple $\boldg =(g_1, \dots, g_n)\in G^n$. This integral can be rewritten as
\begin{equation}\label{eq:renyi_ent_asymm_charged_mom}
    \Delta S_A^{(n)} =  \frac{1}{1-n} \log \slr{ \frac{1}{(\vol G)^{n-1}} \int_{G^n} \d \boldg\,  \Zng \delta \lr{ \prod_{j=1}^n g_j} },
\end{equation}
where $\Zng$ are the (normalized) charged moments of $\rho_A$
\begin{equation}\label{eq:def-Z(g)}
    \Zng =  \frac{\tr(\rho_A U_{A,g_1}  \dots  \rho_A U_{A,g_n})} {\tr(\rho_A^n)}.
\end{equation}

\begin{figure}[t]
\centering \small
\begin{tikzpicture}
  \pgfmathsetmacro{\rdisk}{0.07}
  \pgfmathsetmacro{\hsh}{0.05}
  \pgfmathsetmacro{\vsh}{0.05}

  \draw[] (0,-0.6) -- (4,-0.6) -- (5,0.6) -- (1,0.6) -- cycle;
  
  \draw[] (1.46,-0.05) -- (3.5,-0.05) -- (3.59,0.05) -- (1.55,0.05) -- cycle;

  \draw[white, dashed, very thick] (1.5, 0.6) -- (3.6,0.6); 
  

  \draw [very thick, blue]  (1.46,-0.05) -- node[midway, below]{$g_2$}(3.5,-0.05);
  
  \pgfmathsetmacro{\x}{3.}
    \pgfmathsetmacro{\y}{0.25}
    \pgfmathsetmacro{\r}{0.1}
\draw[dashed, gray] (1.46,-1.55) arc (180:0:\r cm and -\y cm);
\draw[dashed, gray] (3.5,-1.55) arc (180:0:0.125 cm and -\y cm);
    \draw[dashed, gray] (3.747, 0.75) -- (3.747, -1.55);
    \draw[dashed, gray] (1.57, -1.8) -- (3.67, -1.8);
 
   \filldraw[fill=gray, opacity=0.2]
       (3.55,0.7) -- (3.59, 0.05)-- (1.55, 0.05) -- (1.5, 0.7)  arc (180:270:\r cm and -\y cm) -- (3.65, 0.95) arc (270:180:\r cm and -\y cm); ;
  \draw[gray] (3.65,0.95) arc (270:360:\r cm and -\y cm);
  
  \begin{scope}[yshift=-1.5cm]
  \draw[] (0,-0.6) -- (4,-0.6) -- (5,0.6) -- (1,0.6) -- cycle;
  
  \draw[] (1.46,-0.05) -- (3.5,-0.05) -- (3.59,0.05) -- (1.55,0.05) -- cycle;
  \draw[white, dashed, very thick] (1.5, 0.6) -- (3.6,0.6);
    \draw [very thick, blue]  (1.46,-0.05) -- node[midway, below]{$g_1$}(3.5,-0.05);

  \end{scope}
 \filldraw [fill=white, opacity=0.2, dashed] (1.46, -0.05) -- (1.5, -0.6) -- (3.53, -0.6)  -- (3.5, -.05) --cycle ;
  
  \filldraw [fill=gray, opacity=0.2] (1.5, -0.6) -- (1.55, -1.45) -- (3.59, -1.45)  -- (3.53, -0.6) --cycle ;

\end{tikzpicture}
\caption{Riemann surface $\mathcal{M}_n$ for $n=2$ (two sheets) with line defects (in blue) inserted along the branch cut of each sheet. The defects are associated respectively with the group elements $g_1$ and $g_2$. 
The quotient~\eqref{eq:Z-ratio-partition-functions} of the partition functions on this surface with and without the line defects gives the normalized charged moment $\mathcal{Z}_2(\boldg)$, defined in \cref{eq:def-Z(g)}. The Dirac delta in \cref{eq:renyi_ent_asymm_charged_mom} will set $g_2=g_1^{-1}$.}
\label{fig:Riemann_surface}
\end{figure}
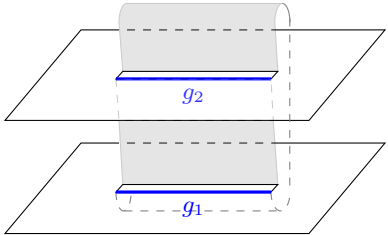

\subsubsection{Interpretation in terms of defects}
In a (1+1) quantum field theory, using the path integral representation of the reduced density matrix $\rho_A$, the neutral moments ${\tr}(\rho_A^n)$ can be identified with the partition
function on an $n$-sheet Riemann surface $\mathcal{M}_n$~\cite{cc-04}. 
If we consider the ground state $\ket{0}$ of the theory, i.e. $\rho=\ket{0}\bra{0}$, and a single interval of length $\ell$ as subsystem $A$, the surface $\mathcal{M}_n$ is constructed as follows. We take the spacetime $\mathcal{M}$ where the theory is defined, which is the complex plane when working in Euclidean time and in the thermodynamic limit (infinite spatial direction). To obtain $\mathcal{M}_n$, we perform a cut on $\mathcal{M}$ along the interval $A=[0,\ell]$, we replicate $n$ times this cut plane, and we sew the copies together along the cuts in a cyclical way, as we show in Fig.~\ref{fig:Riemann_surface} for $n=2$. Denoting as $Z(\mathcal{M}_n)$
the partition function on this surface, the neutral moments of $\rho_A$ are given by
\begin{equation}
    \tr (\rho_A^n)=\frac{Z(\mathcal{M}_n)}{Z(\mathcal{M})^n}.
\end{equation} 

Following the discussion in Sec.~\ref{sec:symm_top_def}, the 
insertion of the operators $U_{A, g_j}$ in this trace, as  in 
Eq.~\eqref{eq:def-Z(g)}, corresponds to putting a defect line along 
the branch cut $[0,\ell]$ of each sheet of $\mathcal{M}_n$ with a 
gluing condition~\eqref{eq:defect-path-integral}, being $g=g_j$, as 
depicted in Fig.~\ref{fig:Riemann_surface}.
If $Z(\mathcal{M}_n^{\boldg})$ stands for the partition function on 
the surface $\mathcal{M}_n$ with these $n$ defect lines, then we have that
\begin{equation}
 \tr(\rho_A U_{A, g_1}\cdots \rho_A U_{A, g_n})=\frac{Z(\mathcal{M}_n^{\boldg})}{Z(\mathcal{M})^n}.
\end{equation}
Therefore, in the ground state, the normalized charged moments $\mathcal{Z}_n(\boldg)$ 
introduced in Eq.~\eqref{eq:def-Z(g)} are the ratio of the partition functions on the 
surface $\mathcal{M}_n$ with and without $n$ defect lines inserted at the branch cut of each sheet,
\begin{equation}\label{eq:Z-ratio-partition-functions}
    \mathcal Z_n(\boldg) = \frac{Z(\mathcal M_{n}^\boldg)}{Z \lr{ \mathcal M_{n} }}.
\end{equation}

If $\rho_A$ is symmetric under $G$, then $[\rho_A, U_{A, g}]=0$ for all $g\in G$. 
As we have previously seen, this implies that the defect lines associated with the insertions $U_{A, g_j}$ are topological and they
can be moved between the sheets of $\mathcal{M}_n$ under continuous 
transformations leaving the partition function $Z(\mathcal{M}_n^{\boldg})$ invariant. In that case, it is possible to fuse them in the same 
sheet, which is equivalent to the equality ${\tr}(\rho_A U_{A, g_1}\dots \rho_A U_{A, g_n})={\tr}(\rho_A^n U_{A, g_1\cdots g_n})$. Since the Dirac delta in \eqref{eq:renyi_ent_asymm_charged_mom} forces the product of all the group elements $g_j$ to be the identity, the fusion yields $U_{A,g_1, \dots g_n} = \id$. Consequently, $\mathcal{Z}_n(\boldg)=1$ and, according to Eq.~\eqref{eq:renyi_ent_asymm_charged_mom}, the R\'enyi entanglement asymmetry vanishes. On the other hand, if $\rho_A$ is not symmetric, $[\rho_A, U_{A, g}]\neq 0$, then the defect lines associated to $U_{A, g_j}$ are not topological. In that case, any continuous deformation of them does change the partition function $Z(\mathcal{M}_n^{\boldg})$ and, as a result, $\mathcal{Z}_n(\boldg) \neq 1$. In this sense, the {\it entanglement asymmetry quantifies how much the defect lines associated with a group are non topological}.

From generic scaling arguments, we can determine the asymptotic behavior of the partition functions $Z(\mathcal{M}_n)$ and $Z(\mathcal{M}_n^{\boldg})$. In two dimensions, the leading order contributions to the free energy $-\log Z$ are proportional to the area $|\mathcal M_n|$ of the surface $\mathcal M_n$, on which the partition function $Z$ is defined. [Of course, strictly speaking, the area $|\mathcal M_n|$ is infinity, but it can be regularized, for instance by imposing periodic boundary conditions both spatial and imaginary time directions for each sheet of $\mathcal{M}_n$, far away from the interval.] Therefore, in the absence of defects,
\begin{equation}\label{eq:free-energy-wo-def}
    -\log Z(\mathcal M_n) = f_\mathrm{bulk} |\mathcal M_n|  + O(1),
\end{equation}
where $f_\mathrm{bulk}$ is the bulk free energy density. In the presence of 
defects, we expect that each of them contributes with an additional
term proportional to the volume of the defect, which in this case is
the length $\ell$ of the interval $A$. The free energy in that case is
\begin{equation}\label{eq:free-energy-w-def}
    - \log Z(\mathcal M_n^\boldg) = f_\mathrm{bulk} |\mathcal M_n| 
    + T_n(\boldg) \ell +  O(1), \quad \text{with} \quad  T_n(\boldg) \coloneqq \sum_{j=1}^n t(g_j),
\end{equation}
and $t(g_j)$ can be interpreted as the line tension of the defect associated with the insertion $U_{A, g_j}$. All these terms are cut-off dependent and, therefore, non universal. Plugging Eqs.~\eqref{eq:free-energy-wo-def}-\eqref{eq:free-energy-w-def} into Eq.~\eqref{eq:Z-ratio-partition-functions}, one sees that the bulk contribution in the free energy cancels, and the charged moments
$\mathcal{Z}_n(\boldg)$ decay exponentially with the subsystem length $\ell$ as 
\begin{equation}\label{Z(g)-exp-tension}
    \mathcal Z_n(\boldg) = e^{- T_n(\boldg) \ell + O(1)}.
\end{equation}

If the theory is critical, the conical singularities at the branch 
points of the surface $\mathcal{M}_n$ give rise in 
Eqs.~\eqref{eq:free-energy-wo-def} and~\eqref{eq:free-energy-w-def} 
to an extra universal (cut-off independent) term,
\begin{equation}
 -\log Z(\mathcal{M}_n^{\boldg})= f_\mathrm{bulk} |\mathcal{M}_n|  
    + T_n(\boldg) \ell  -\log Z_{\rm CFT}(\mathcal{M}_n^{\boldg})+O(1),
\end{equation}
which, as argued for instance in Ref.~\cite{CardyPeschel:1988finite}, behaves as $-\log Z_{\rm CFT}(\mathcal{M}_n^{\boldg})\propto \log\ell$. 
The presence of defect lines may in general modify the coefficient of this term, so 
\begin{equation}
 \frac{Z_{\rm CFT}(\mathcal{M}_n^{\boldg})}{Z_{\rm CFT}(\mathcal{M}_n)}=\ell^{-\beta_n(\boldg)}
\end{equation}
and it does not cancel in the ratio~\eqref{eq:Z-ratio-partition-functions} of partition functions that gives the normalized charge 
moments $\mathcal{Z}_n(\boldg)$. Therefore, for a critical system, 
we expect
\begin{equation}\label{eq:Zng-critical}
    \Zng = e^{-T_n(\boldg)\ell+O(1)} \ell^{-\beta_n(\boldg)},
\end{equation}
where the coefficient $\beta_n(\boldg)$ is universal and can be computed in the infrared (IR) CFT that describes the critical system. It depends on the specific CFT and the nature of the defects corresponding to the group $G$ under study, and we do not have a generic expression for it. Its computation has to be worked out case by case. In this paper, we calculate it in the massless Majorana fermion field theory for a $U(1)$ group
for which the defects are marginal.

\subsubsection{Asymptotic behavior}\label{sec:generic_asymp_beh}

Before delving into the study of the charged moments and entanglement asymmetry in a particular theory, it is  insightful to explore the implications of the generic result of~\cref{eq:Zng-critical} for the asymptotic behavior of the entanglement asymmetry in the limit of large subsystem size $\ell$.

When we plug Eq.~\eqref{eq:Zng-critical} in Eq.~\eqref{eq:renyi_ent_asymm_charged_mom}, we have to perform an $n$-fold integral over the group $G$. 
Since the leading term in Eq.~\eqref{eq:Zng-critical} decays exponentially with $\ell$, the main contribution to this integral comes from the points $\boldh\in G^{n}$ where $\mathcal{Z}_n(\boldh)=1$ (i.e. where both $T_n(\boldh)$ and $\beta_n(\boldh)$ vanish). 
These correspond to the elements of $G$ that leave the reduced density matrix $\rho_A$ invariant and form a symmetry subgroup $H$ of $G$. i.e.
\begin{equation}
    H = \clr{ h \in G \mid U_{A,h} \rho_A U_{A,h}^\dagger = \rho_A }.
\end{equation}
Therefore, the strategy is to perform a saddle point approximation of the integral~\eqref{eq:renyi_ent_asymm_charged_mom} around
the points $\boldh\in H^n$; see also Refs.~\cite{CapizziVitaleMPS:2023, Murciano:asymmetryXY, bkccr-23, amvc-23}.

For simplicity, let us assume that $H$ is a finite subgroup. 
In the integral \eqref{eq:renyi_ent_asymm_charged_mom_no_dirac_delta} the numerator $\tr(U_{A,g_1} \rho_A U_{A,g_1^{-1}} \dots U_{A,g_n} \rho_A U_{A,g_n^{-1}})$ is invariant under a right multiplication $(g_1, \dots, g_n) \mapsto (g_1 h_1, \dots, g_n h_n)$. Consequently, all the saddle points $\boldh\in H^{n}$ contribute equally. Then, to calculate the integral~\eqref{eq:renyi_ent_asymm_charged_mom} for $\ell\gg 1$, we can expand it around the identity point $(\mathrm{Id}, \dots, \mathrm{Id}) \in G^n$, where $\mathrm{Id}$ is the identity in $G$, and multiply the result by the total number of saddle points, which is given in terms of the cardinality  $|H|$ of $H$ as $|H|^{n-1}$. 
We finally perform the integral by choosing some local coordinates on the group around the identity.

In a neighborhood $\mathcal{U}_\mathrm{Id}\subset G$ of the identity, the group elements $g$ can be written as $g=e^{iX}$, where $X$ is an element of the
Lie algebra $\mathfrak{g}$ associated with $G$, of dimension $d=\dim G$. Let $\{J_a\}$, $a=1, \dots, d$, be generators of $\mathfrak{g}$, if we take the local coordinate chart $\mathrm{x}=(x_1, \dots, x_d)\in\mathbb{R}^d\mapsto g(\mathrm{x})=e^{i\sum_a x_a J_a}$, then, for an arbitrary function $f(g)$ on $G$, we have
\begin{equation}
    \int_{\mathcal U_\mathrm{Id}} \d g \, f(g) = \int_{g^{-1}(\mathcal U_\mathrm{Id})} \mu(\mathrm{x}) \d \mathrm{x} \,  f(g(\mathrm{x})), 
\end{equation}
where $\mu(\mathrm{x}) \d x$ is the Haar measure of $G$ in the local coordinates $\mathrm{x}$. Since we have to perform an $n$-fold integral over $G$, we denote by $\bx$ the coordinates for $G^n$, that is $\bx = (\mathrm{x}_1, \dots, \mathrm{x}_n) \in \RR^{dn}$. 
Now we can express the exponents $T_n(\boldg)$ and $\beta_n(\boldg)$ of the charged moments~\eqref{eq:Zng-critical} in coordinates and expand them around the identity, which corresponds to $\bx = \mathbf{0}$,
\begin{equation}
\begin{aligned}
    T_n(\boldg(\bx)) &= \frac{1}{2} \bx^{\mathrm{t}} \hessTn \bx+ O(\bx^3), \\
    \beta_n(\boldg(\bx)) &= \frac{1}{2} \bx^\mathrm t \hessbetan \bx + O(\bx^3),
\end{aligned}
\end{equation}
where $\hessTn$ and $\hessbetan$ are $dn\times dn$ Hessian matrices, made of $n \times n$  blocks of dimension $d \times d$. Therefore, in the local coordinate chart that we are considering, for large $\ell$ the $n$-fold integral~\eqref{eq:renyi_ent_asymm_charged_mom} reads 
\begin{equation}\label{eq:int_asymp_beh}
    \begin{aligned}
        \int_{G^n} \Zng \delta \Big( \prod_{j} g_j\Big) \d \boldg \sim |H|^{n-1}\mu(0)^{n-1} \int_{\RR^{dn}}  \d \bx \, e^{-\frac{1}{2} \bx^{\mathrm{t}} (\hessTn \ell + \hessbetan \log \ell) \bx } \delta \lr{ \sum_{j=1}^n \mathrm{x}_j } .
    \end{aligned}
\end{equation}
Here the factor $|H|^{n-1}$ counts the total number of saddle points. 
In coordinates, the Dirac delta $\delta\left(\prod_j g_j\right)$ over 
the group $G$ is replaced by $\delta\left(\sum_{j=1}^x \mathrm{x}_j\right)/\mu(0)$. Notice that we have also expanded the measure 
$\mu(\mathrm{x})$ around $\mathrm{x}=0$ and restricted to the zeroth 
order term $\mu(0)$ since the next order terms yield subleading 
corrections in $\ell$.

Since $T_n(\boldg)$ is the sum of the contributions of each defect line according
to Eq.~\eqref{eq:free-energy-w-def}, $\hessTn$ is block diagonal, $\hessTn=\id_n \otimes \hess_t$, where $\hess_t$ is the $d \times d$
dimensional Hessian of $t(g(x))$. 
Due to the cyclic property of the trace, the coefficient $\beta_n(\boldg)$ is symmetric under cyclic permutations of the entries $g_j$ of $\boldg$. Thus $\hessbetan$ is a block-circulant matrix; that is, it has the block structure
\begin{equation}
    \hessbetan = 
\begin{pmatrix}
\mathsf C_1      & \mathsf C_{n} & \cdots  & \mathsf C_3     & \mathsf C_2     \\
\mathsf C_2      & \mathsf C_1     & \mathsf C_{n} &         & \mathsf C_3     \\
\vdots   & \mathsf C_2     & \mathsf C_1     & \ddots  & \vdots  \\
\mathsf C_{n-1}  &         & \ddots  & \ddots  & \mathsf C_{n} \\
\mathsf C_{n}  & \mathsf C_{n-1} & \cdots  & \mathsf C_2     & \mathsf C_1     \\
\end{pmatrix},
\end{equation}
with blocks $\mathsf C_j$ of size $d \times d$. A block-circulant matrix can be diagonalized in blocks $\mathsf D_p$, $p=0,\dots,n-1$, with a Fourier transform of the blocks $\mathsf C_j$,
\begin{equation}\label{eq:D_p}
    \mathsf D_p = \sum_{j=1}^{n} \mathsf C_j e^{i \frac{2\pi}{n} jp} .
\end{equation}
Therefore, if we apply the change of variables
\begin{equation}
    {\rm x}_j = \frac{1}{\sqrt n} \sum_{p=0}^{n-1} \omega_p e^{-i\frac{2\pi}{n} jp}, \quad  \qquad j= 1,\dots, n, 
\end{equation}
then the integral~\eqref{eq:int_asymp_beh} becomes
\begin{equation}
    \int_{G^n} \Zng \delta \Big( \prod_{j} g_j\Big) \d \boldg  \sim \frac{|H|^{n-1} \mu(0)^{n-1}}{\sqrt n} \int_{\RR^{dn}} \d \boldsymbol{\omega} e^{-\frac{1}{2} \sum_{p=0}^{n-1} \omega_p ( \mathsf H_t  \ell +  \mathsf D_p  \log \ell) \omega_p } \delta \lr{\omega_0 }.  
\end{equation}    
Integrating out the variable $\omega_0$, we find
\begin{equation}\label{eq:int_group_1}
    \int_{G^n} \Zng \delta \Big( \prod_{j} g_j\Big) \d \boldg\sim \frac{|H|^{n-1} \mu(0)^{n-1}}{\sqrt n} \prod_{p=1}^{n-1} \,  \int_{\RR^{d}}  \d \omega \, e^{-\frac{1}{2} \omega ( \mathsf H_t  \ell +  \mathsf D_p  \log \ell) \omega }  .
\end{equation}
The remaining integral is Gaussian and we can easily perform it, if we assume that $\mathsf H_t\ell+\mathsf D_p \log\ell$ 
is a symmetric definite-positive matrix,
\begin{equation}
     \int_{\RR^{d}} \d \omega e^{-\frac{1}{2} \omega ( \mathsf H_t  \ell +  \mathsf D_p  \log \ell) \omega}  = \lr{ \det \lr{2\pi(\hess_t \ell + \mathsf D_p \log \ell}^{-1} }^{1/2}.
\end{equation}
Plugging it in Eq.~\eqref{eq:int_group_1}, we obtain
\begin{equation}\label{eq:int_group_2}
 \int_{G^n} \Zng \delta \Big( \prod_{j} g_j\Big) \d \boldg \sim \frac{(2\pi\ell)^{\frac{d(1-n)}{2}}}{\sqrt{n}}\left(\frac{|H|\mu(0)}{\sqrt{\det \mathsf H_t}}\right)^{n-1}\prod_{p=0}^{n-1}\left(\det \lr{\id + \hess_t^{-1}\mathsf D_p \frac{\log \ell}{\ell} }\right)^{1/2}.
\end{equation}
This result is independent of the local coordinate chart that we consider to perform the integration.
In fact, under a change of local coordinates $\mathrm{x}\mapsto \mathrm{y}(\mathrm{x})$, the measure $\mu(\mathrm{x})$ transforms 
as $\mu(\mathrm{x})=\left|\det(\partial y^{\sigma}/\partial x^b)\right|\mu'(\mathrm{y})$ and, because quadratic forms are
$(0,2)$-tensor fields, the determinant of the Hessian $\mathsf H_t$ transforms as $\det \mathsf{H}_t(\mathrm{x})= \left|\det(\partial y^{\sigma}/\partial x^b)\right|^2\det \mathsf{H}_t'(\mathrm{y})$. Therefore, the quotient $\mu(0)/\sqrt{\det\mathsf{H}_t}$ is coordinate
independent. The same applies for the terms $\det(\id+\mathsf H_t^{-1}\mathsf D_p \log\ell/\ell)$.

Finally, applying~\eqref{eq:int_group_2} in Eq.~\eqref{eq:renyi_ent_asymm_charged_mom} and using the identity $\log \det M= \tr \log M$ for a matrix $M$,
we find that the R\'enyi entanglement asymmetry for a compact Lie group $G$ in the ground state of a critical 
one dimensional quantum system behaves as
\begin{equation}\label{eq:renyi_ent_asymm_generic}
 \Delta S_A^{(n)}=\frac{\dim G}{2}\log\ell+a_n+b_n\frac{\log\ell}{\ell}+\cdots,
\end{equation}
where 
\begin{equation}
 a_n= \log \frac{\vol G}{|H|} + \frac{1}{2} \log  \frac{n^{\frac{1}{n-1}} \det \mathsf H_t }{(2\pi)^{\dim G} \mu(0)^2},
\end{equation}
and 
\begin{equation}
 b_n=\frac{1}{2(n-1)} \sum_{p=1}^{n-1} \tr\left(\mathsf D_p\mathsf H_t^{-1}\right).
\end{equation}
Eq.~(\ref{eq:renyi_ent_asymm_generic}) is the first main result of this paper. We stress that the first two terms in (\ref{eq:renyi_ent_asymm_generic}), of order $O(\log \ell)$ and $O(1)$ respectively, have already been observed in the XY spin chain when considering the particle number $U(1)$ (a)symmetry~\cite{Murciano:asymmetryXY}, and more generically for matrix product states in Ref.~\cite{CapizziVitaleMPS:2023}. Crucially, what is new here is the last term in Eq.~(\ref{eq:renyi_ent_asymm_generic}), of order $O(\log \ell/\ell)$. While the terms of order $O(\log \ell)$ and $O(1)$ are present in the ground state of critical and non-critical systems alike, the term of order $O(\log \ell/\ell)$ only appears when the system is at a critical point.

This $\log\ell/\ell$ term appears only when the
exponent $\beta_n(\boldg)$ in Eq.~(\ref{eq:Zng-critical}) is non-zero. Although this exponent is universal, the coefficient $b_n$ is non-universal since it also depends on the defect tension $T_n(\boldg)$ (via $\mathsf H_t$), which is cut-off dependent. Semi-universal corrections of the form $\log\ell/\ell$ have been found in, e.g., the corner free energy in critical systems~\cite{sd-13} and in the 
ground state full counting statistics of the critical XY spin chain~\cite{Stephan:2013logarithmic}.

We finally discuss the group structures that were not considered earlier.  
When both $G$ and $H$ are finite, it is straightforward to show that the logarithmic term is vanishing, the $O(1)$ term is just $\log (|G|/|H|)$ and there are no $\log\ell/\ell$ corrections (see also \cite{CapizziVitaleMPS:2023}). 
When both $G$ and $H$ are continuous, the leading $\log\ell$ term has a prefactor equal to $(\dim G-\dim H)/2$, but the explicit expressions for the subleading terms are more cumbersome and not very illuminating.

\section{The XY spin chain and the massless Majorana fermion field theory}\label{sec:setup}

In the rest of the paper, we focus on a particular gapless 
system: the XY spin chain at the Ising critical line. We consider its ground state, and compute 
the charged moments and the R\'enyi entanglement asymmetry associated
with the rotations of the spin around the $z$-axis.

The Hamiltonian of the XY spin chain is 
\begin{equation}\label{eq:XY_ham}
   H_\mathrm{XY} = - \frac{1}{2} \sum_{j \in \ZZ} \lr{ \frac{1+\gamma}{2} \sigma^x_j \sigma^x_{j+1} + \frac{1-\gamma}{2} \sigma^y_j \sigma^y_{j+1} + h \sigma^z_j },
\end{equation} 
where $\sigma_j^\alpha$ are the Pauli matrices at the site $j$.
The parameter $\gamma$ tunes the anisotropy between the couplings in
the $x$ and $y$ components of the spin and $h$ is the strength  of 
the transverse magnetic field. The XY spin chain is gapless
along the lines $\gamma=0$, $|h|<1$ and $\gamma\neq 0$, $|h|=1$ in parameter space, and the scaling limits along those lines are respectively the massless Dirac and Majorana fermion field theories.

For $\gamma\neq 0$, the Hamiltonian of Eq.~\eqref{eq:XY_ham} is not invariant under the rotations $U_{\alpha}=e^{i\alpha Q}$ around the $z$-axis, generated by the transverse magnetization
\begin{equation}\label{eq:transv_mag}
Q=\frac{1}{2}\sum_{j\in\ZZ}(\sigma_j^z-\II),
\end{equation}
except for $\alpha=\pi$, which corresponds to the $\ZZ_2$ spin flip symmetry. The entanglement 
asymmetry associated with this $U(1)$ symmetry has been thoroughly 
studied in Ref.~\cite{Murciano:asymmetryXY} for the ground 
state of~\eqref{eq:XY_ham} outside the critical lines $\gamma\neq 0$, $|h|=1$ using exact methods on the lattice. In that case, the charged moments $\mathcal{Z}_n(\boldsymbol{\alpha})$ decay exponentially for large subsystem size $\ell$ as in Eq.~\eqref{Z(g)-exp-tension}, where the coefficient $T_n(\boldsymbol{\alpha})=\sum_{j=1}^n t(\alpha_j)$ is the sum of the string tension $t(\alpha_j)$ of each defect, which here is given by~\cite{Murciano:asymmetryXY}
\begin{equation}\label{eq:t_n_xy}
t(\alpha)=-\int_{-\pi}^\pi\frac{\d k}{4\pi}\log(i\cos\xi_k\sin\alpha+\cos\alpha),
\end{equation}
with 
\begin{equation}\label{eq:bogo_angle}
e^{i\xi_k}=\frac{h-\cos k+i\gamma \sin k}{\sqrt{(h-\cos k)^2+\gamma^2\sin^2k}}.
\end{equation}
Note that the string tension $t(\alpha)$ is not real; as we will see in Sec.~\ref{sec:CFT_calculation_Delta}, this is related to the fact that the gluing conditions of the defects associated with this $U(1)$ group make the theory non-Hermitian.

To obtain the R\'enyi entanglement asymmetry, we can apply the general result of Eq.~\eqref{eq:renyi_ent_asymm_generic}. In this case, since $G=U(1)$, we have that
$\dim G=1$, $\vol G=2\pi$, $\mu(0)=1$, and the symmetric subgroup is 
the $\mathbb{Z}_2$ spin-flip symmetry, $H=\mathbb{Z}_2$. Since $\dim G=1$, the block $\hess_t$ is a scalar and it is given by Eq.~\eqref{eq:t_n_xy} such that $\hess_t=t''(0)=\partial^2 t(\alpha)/\partial \alpha^2|_{\alpha=0}$.

In Ref.~\cite{Murciano:asymmetryXY}, the following result was derived for the XY chain out of the critical line $|h| = 1$: 
\begin{equation}
    \Delta S_A^{(n)} = \frac{1}{2} \log \ell + 
\frac{1}{2} \log \frac{\pi t''(0) n^{1 /(n-1)}}{2} + O(\ell^{-1}),
\end{equation}
with 
\begin{equation}\label{eq:hess_t_xy}
t''(0)= \begin{cases}
\frac{1}{2}\frac{\gamma}{\gamma+1}, & |h| < 1, \\
\frac{1}{2}\frac{\gamma^2}{1-\gamma^2}\left(\frac{|h|}{\sqrt{h^2+\gamma^2-1}}-1\right), & |h|>1 .
\end{cases}
\end{equation}
Notice that $t''(0)$ is continuous at $|h| = 1$, reflecting the fact that this result also applies along the critical line. Indeed, in this case, the string tension $T_n(\boldsymbol{\alpha})$ is still given by Eq.~\eqref{eq:t_n_xy}, and following the same steps as in Ref.~\cite{Murciano:asymmetryXY} one arrives at the same result.

However, crucially for this paper, along the critical line $\gamma\neq 0$, $|h|=1$, we also expect that the charged moments $\mathcal{Z}_n(\boldsymbol{\alpha})$ contain the
algebraically decaying factor of Eq.~\eqref{eq:Zng-critical}, according to the general reasoning of \cref{sec:generic_asymp_beh}. However, an analytical expression for the coefficient $\beta_n(\aalpha)$ is unknown. In what follows, we will obtain it by exploiting the conformal invariance of the underlying field theory. 

As we have already mentioned, the scaling limit of the XY spin chain~\eqref{eq:XY_ham}
along the critical lines $\gamma\neq 0$, $|h|=1$ is the massless Majorana fermion field theory, whose Hamiltonian is
\begin{equation}\label{eq:majorana_ham}
    H=\frac{1}{2 i} \int_{\RR} \psi(x) \partial_x \psi(x) -\bar\psi(x) \partial_x \bar\psi(x) \d x,
\end{equation}
where the Majorana fields $\psi(x)$ and $\bar \psi(x)$ satisfy the 
algebra 
\begin{align}
    \{\psi(x), \psi(y)\} &=  \delta(x-y), & \psi^\dagger(x) &= \psi(x),  \\
    \{\bar \psi(x), \bar\psi(y)\} &=  \delta(x-y),& \bar \psi^\dagger(x) &= \bar \psi(x),
\end{align}
and $\{\psi(x), \bar \psi(y)\} = 0 $. The $U(1)$ charge operator in
Eq.~\eqref{eq:transv_mag} corresponds in this field theory to 
\begin{equation}\label{eq:charge_majorana_ft}
    Q = i \int_\RR \psi(x) \bar\psi(x) \d x .
\end{equation}
The details on the derivation of the Hamiltonian~\eqref{eq:XY_ham}
and $Q$ in the continuum limit are reported in Appendix~\ref{app:continuum-limit-majorana}.

The transformations 
\begin{equation}
    \label{eq:U-alpha-majorana}
    U_{A,\alpha} = e^{i \alpha Q_A} = \exp \lr{- \alpha \int_A  \psi(x) \bar\psi(x) \d x}
\end{equation}
generated by the charge~\eqref{eq:charge_majorana_ft} in a subsystem $A$ act on the fields 
$\psi(x)$, $\bar \psi(x)$ in the following way 
\begin{equation}\label{eq:gluing-horizontal-defect}
    U_{A,\alpha}^\dagger 
    \Psi(x) U_{A,\alpha} = 
    \tilde{R}_\alpha
    \Psi(x),  \qquad \text{if } x \in A, 
\end{equation} 
with
\begin{align}
    \label{eq:Ralpha}  
    \Psi = \begin{pmatrix}
        \psi \\ \bar \psi
    \end{pmatrix},
    \quad 
    \tilde{R}_\alpha = \begin{pmatrix}
        \cos \alpha & -\sin \alpha \\
        \sin \alpha & \cos \alpha
    \end{pmatrix} \in SO(2).
\end{align}
The group action consists of a rotation that mixes $\psi$ and 
$\bar \psi$. In general, this is not a symmetry of the theory,
unless $\alpha=\pi$, for which $\psi\mapsto -\psi$ and $\bar \psi\mapsto -\bar \psi$. For $\alpha$ purely imaginary, the defect can be realized in the classical 2d Ising model by rescaling the couplings on all the bonds that intersect the defect line. A dictionary between the two realizations is given, e.g., in \cite{bb-15}. 

Crucially for our analysis, the field $\psi(x)\bar \psi(x)$ has 
scaling dimension 1, therefore the line defect implemented by 
$U_{A, \alpha}$ corresponds to a marginal perturbation of the CFT action along the line. 
This is very important for the calculations reported in Section~\ref{sec:CFT_calculation_Delta}, as it it introduces a non-trivial dependence of the CFT partition function depend on the defect strength $\alpha$. Indeed, if the perturbation were instead irrelevant, then the effects of the defect would be renormalized to zero in the IR limit. If the perturbation were relevant, then the defect would flow to some fixed point in the IR, corresponding to some boundary condition along the line, and the CFT partition function would also be independent on the precise value of $\alpha$. For instance, such a situation would occur if we looked at the asymmetry with respect to rotations around the $x$-axis, as opposed to the $z$-axis, corresponding to replacing $\sigma^z_j$ with $\sigma_j^x$ in Eq.~(\ref{eq:transv_mag}). In the CFT, this would then correspond to a perturbation by the relevant operator $\sigma(x)$ with scaling dimension $1/8$. The defect line would flow to a fixed boundary condition in the IR, and this would completely change the way we do the analysis, see in particular Ref.~\cite{lamacraft2008order} for more details on that situation.

\section{Calculation of the scaling dimension associated to \texorpdfstring{$n$}{n} defects in the Majorana CFT}
\label{sec:CFT_calculation_Delta}

In Sec.~\ref{sec:asymm_def}, we have seen that the charged moments $\mathcal{Z}_n(\aalpha)$ can be cast as the ratio $Z(\mathcal{M}_n^{\aalpha})/Z(\mathcal{M}_n)$ between the partition function $Z(\mathcal{M}_n^{\aalpha})$ of the model on the $n$-sheet Riemann surface $\mathcal{M}_n$ with $n$ defect lines with strengths ${\aalpha} = ( \alpha_1, \dots, \alpha_n )$ along its branch cuts and the one, $\mathcal{Z}(\mathcal{M}_n)$, without them. As we discussed, in critical systems, this ratio contains a universal term $Z_{\rm CFT}(\mathcal{M}_n^{\aalpha})/Z_{\rm CFT}(\mathcal{M}_n)$, fully determined by the CFT that describes the low-energy physics. 
In this section, we study it in the massless Majorana fermion theory~\eqref{eq:majorana_ham} for the marginal defect lines~\eqref{eq:U-alpha-majorana}.

When there are no defects, it is well-known~\cite{hlw-94,cc-04,cc-09} that, for a generic CFT,
\begin{equation}\label{eq:Z_CFT_no_defects}
    \frac{Z_\mathrm{CFT}(\mathcal M_n)}{Z_\mathrm{CFT} (\mathcal M)^n} \propto \ell^{-2 \delta_n} , \qquad \delta_n = \frac{c}{12} \left( n-\frac{1}{n}\right) ,
\end{equation}
where $c$ is the central charge of the CFT, which for the massless 
Majorana fermion is $c=1/2$. 

In the massless Majorana fermion theory, when we insert the $n$ marginal defect lines along each branch cut of the surface $\mathcal{M}_n$,
the result~\eqref{eq:Z_CFT_no_defects} changes as 
\begin{equation}
    \label{eq:Delta_alpha}
    \frac{Z_{\rm CFT}(\mathcal M_n^{{\aalpha}})}{Z_{\rm CFT}(\mathcal{M})^n} \propto \ell^{ - 2\left(\delta_n +\frac{\tilde{\Delta}_n(\aalpha)}{n}\right)} ,
\end{equation}
as we will show below. The contribution of the $n$ marginal defects is encoded in the exponent $\tilde \Delta_n(\aalpha)$. Then the ratio of the partition functions on the surface $\mathcal M_n$ with and without defects is 
\begin{equation} \label{eq:ratio-partition-functions-cft}
    \frac{Z_\mathrm{CFT}(\mathcal M_n^{\aalpha})}{Z_\mathrm{CFT}(\mathcal M_n)} = \ell^{- \frac{2}{n} \tilde \Delta_n(\aalpha)},
\end{equation}
and, comparing with Eq.~\eqref{eq:Zng-critical}, $\beta_n(\aalpha) = \frac{2}{n} \tilde \Delta_n(\aalpha)$. The rest of the section will be devoted to deriving Eq.~\eqref{eq:Delta_alpha} and computing explicitly the coefficient $\tilde{\Delta}_n(\boldsymbol{\alpha})$.

\subsection{Conformal mapping to the cylinder with \texorpdfstring{$n$}{n} defect lines}

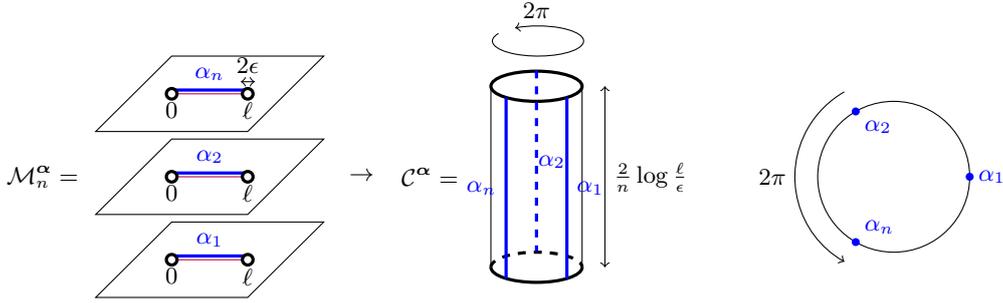
\begin{figure}[t]
\centering \small

\begin{tikzpicture}

\begin{scope}[xshift=-5cm]
  \pgfmathsetmacro{\rdisk}{0.07}
  \pgfmathsetmacro{\hsh}{0.05}
  \pgfmathsetmacro{\vsh}{0.05}

  \draw[] (0,-0.5) -- (2,-0.5) -- (3,0.5) -- (1,0.5) -- cycle;
  
    \draw [very thick, blue]  (1+\hsh,\vsh) -- node[midway, above]{$\alpha_n$}  (2-\hsh,\vsh);
    \draw[purple]  (1+\rdisk,0) --  (2-\rdisk,0);
    \draw[very thick] (1,0) circle (\rdisk);
    \draw[very thick] (2,0) circle (\rdisk);
    \draw[<->] (1.91,0.16) --  (1.91+2.6*\rdisk,0.16) node[pos=0.5,above]{\small $2\epsilon$};
  \node[below] at (1,0.) {$0$};
  \node[below] at (2,0.) {$\ell$};
  
  \node at (3.5,-1.1) {$\to$};
  \node at (-0.7, -1.1) {$\mathcal M_n^{\aalpha} = $};
  \node at (4.4, -1.1) {$\mathcal C^{\aalpha}=$};

\begin{scope}[yshift=-1.1cm]
  \draw[] (0,-0.5) -- (2,-0.5) -- (3,0.5) -- (1,0.5) -- cycle;
  
    \draw [very thick, blue]  (1+\hsh,\vsh) -- node[midway, above]{$\alpha_2$}  (2-\hsh,\vsh);
    \draw[purple]  (1+\rdisk,0) --  (2-\rdisk,0);
    \draw[very thick] (1,0) circle (\rdisk);
    \draw[very thick] (2,0) circle (\rdisk);
  \node[below] at (1,0.) {$0$};
  \node[below] at (2,0.) {$\ell$};
\end{scope}

\begin{scope}[yshift=-2.2cm]
  \draw[] (0,-0.5) -- (2,-0.5) -- (3,0.5) -- (1,0.5) -- cycle;
  
    \draw [very thick, blue]  (1+\hsh,\vsh) -- node[midway, above]{$\alpha_1$}  (2-\hsh,\vsh);
    \draw[purple]  (1+\rdisk,0) --  (2-\rdisk,0);
    \draw[very thick] (1,0) circle (\rdisk);
    \draw[very thick] (2,0) circle (\rdisk);
  \node[below] at (1,0.) {$0$};
  \node[below] at (2,0.) {$\ell$};
\end{scope}

\begin{scope}[xshift=5.8cm,yshift=-1.1cm]
    \pgfmathsetmacro{\x}{0.6}
    \pgfmathsetmacro{\y}{1.2}
    \pgfmathsetmacro{\r}{0.2}
    \pgfmathsetmacro{\vsh}{0.05}
    \pgfmathsetmacro{\hsh}{0.4}
    \pgfmathsetmacro{\vshh}{0.01}
    \pgfmathsetmacro{\hshh}{0.0}
    \pgfmathsetmacro{\vshhh}{0.05}
    \pgfmathsetmacro{\hshhh}{-0.4}
    \draw[very thick] (0,\y) ellipse (\x cm and \r cm);
    
    \draw (-\x,\y) -- (-\x,-\y);
    \draw (\x,\y) -- (\x,-\y);
    
    \draw[very thick, dashed] (-\x,-\y) arc (0:180:-\x cm and \r cm);
    \draw[very thick] (-\x,-\y) arc (0:180:-\x cm and -\r cm);
    
    \draw [very thick, blue]  (\hsh,\y-\r+\vsh) -- node[midway, right]{$\alpha_1$} (\hsh,-\y-\r+\vsh);
    \draw [very thick, dashed, blue]  (\hshh,\y+\r+\vshh) -- (\hshh,-\y+\r+\vshh);
    \draw [very thick, blue]  (\hshhh,\y-\r+\vshhh) -- node[midway, left]{$\alpha_{n}$}(\hshhh,-\y-\r+\vshhh);
    \node[blue] at (0.2, \r) {$\alpha_2$};
    
    \draw[<->] (\x+0.3,-\y) -- node[midway, right]{$\frac{2}{n}\log\frac{\ell}{\epsilon}$} (\x+0.3,\y);
    \draw[->,] (-0.5,\y+0.7) arc (-210:120:\x cm and \r cm);
    \node at (0,\y + 1) {$2\pi$};
  
\end{scope}
\end{scope}
\begin{scope}[xshift=5.5cm, yshift=-1.1cm]
  \coordinate (O) at (0,0); 
  
  \draw (O) circle[radius=1cm];
  
  \foreach \i in {1,...,3} {
    \coordinate (X\i) at ({360/3 * (\i - 1)}:1cm);
  }

  \foreach \i in {1,...,3} {
    \ifnum \i = 1
      \fill[blue] (X\i) circle[radius=1.5pt] node[right] {$\alpha_\i$};
    \else
      \ifnum \i = 2
        \fill[blue] (X\i) circle[radius=1.5pt] node[below right] {$\alpha_\i$};
      \else
        \ifnum \i = 3
          \fill[blue] (X\i) circle[radius=1.5pt] node[above right] {$\alpha_n$};
        \else
          \fill[blue] (X\i) circle[radius=1.5pt];
        \fi
      \fi
    \fi
  }
  
  \draw[<-] (240:1.3cm) arc (240:120:1.3cm) node[midway, left] {$2\pi$};
  \end{scope}
\end{tikzpicture}
\caption{
On the left, we represent the $n$-sheet Riemann surface $\mathcal M_n$ with $n$ marginal defect lines inserted along the branch cut $[0,\ell]$ of each replica sheet, which arises in the calculation of the ground state charged moments $\mathcal{Z}_n(\boldsymbol{\alpha})$. At the branch points $0$ and $\ell$, two disks of radius $\epsilon$ have been removed as UV cut-off. Under the conformal transformation~\eqref{eq:conformal-map}, $\mathcal{M}_n$ is mapped into the cylinder $\mathcal{C}$ on the middle, of circumference $2\pi$ and height $\frac{2}{n} \log \frac{\ell}{\epsilon}$. The defect lines in $\mathcal{M}_n$ are mapped into $n$ evenly spaced vertical defects at the points $x_j=\frac{2\pi j}{n}, j=1, \dots, n$. The CFT partition functions on these two surfaces with the marginal defects are equal. On the right, top view of the cylinder $\mathcal{C}$ with
the defects.}
\label{fig:conformal-maps}
\end{figure}

To determine the partition function $Z_{\mathrm{CFT}}(\mathcal{M}_n^{\aalpha})$ with the $n$ marginal defect lines, we perform the conformal transformation
\begin{equation}\label{eq:conformal-map}
    z \mapsto w(z) = i\log \slr{ \left(\frac{z}{z-\ell}\right)^{1/n} }.
\end{equation} 
If at the branch points $z=0$ and $z=\ell$ of the Riemann surface $\mathcal{M}_n$ we remove a disk of radius $\epsilon$ as a UV cut-off, then Eq.~\eqref{eq:conformal-map} maps $\mathcal{M}_n$ to a cylinder with circumference $2\pi$ and height $W=\frac{2}{n}\log(\ell/\epsilon)$, which we denote as $\mathcal{C}$, see Fig.~\ref{fig:conformal-maps}. We choose as coordinates of the cylinder $w=x+i\tau$, with $x\sim x+2\pi$ and $\tau \in [-\frac{1}{n}\log(\ell/\epsilon), \frac{1}{n}\log(\ell/\epsilon)]$. 

Under Eq.~\eqref{eq:conformal-map}, the $n$ branch cuts $[0, \ell]$ of $\mathcal{M}_n$ are mapped to the equally-spaced lines $x_j=\frac{2\pi j}{n}, \, j=1,\dots, n$ ($x_n=2\pi$ is identified with $x=0$) on the cylinder, as we illustrate in Fig.~\ref{fig:conformal-maps}. Thus, on the cylinder $\mathcal{C}$, the $n$ marginal defects are inserted along these lines. We assume the Majorana fields $\psi, \bar\psi$ to have trivial monodromy on $\mathcal M_n$ along the cycle that connects all the replicas. Therefore, after the map~\eqref{eq:conformal-map}, these fields satisfy anti-periodic boundary conditions on the cylinder since they have half-integer spin~\cite{difrancesco}.

The next step is to carefully determine the gluing condition that is satisfied by the Majorana fields $\psi$ and $\bar{\psi}$ across each defect after the conformal map~\eqref{eq:conformal-map} to the cylinder. In the previous section, we found that, on the Riemann surface $\mathcal{M}_n$, the gluing condition across a defect with strength $\alpha_j$ is given by Eq.~(\ref{eq:gluing-horizontal-defect}), i.e.
\begin{equation}\label{eq:gluing-horizontal-defect-2}
      \Psi (z=x + i 0^+ ) = \tilde{R}_{\alpha_j}  \Psi (z=x + i 0^- ),  \qquad {\rm for} \quad x \in [0, \ell],
\end{equation}
where the $2\times 2$ matrix $\tilde{R}_{\alpha_j}$ is defined in Eq.~(\ref{eq:Ralpha}). Crucially, this gluing condition changes under the conformal transformation~\eqref{eq:conformal-map}. Indeed, since the Majorana fields $\psi$ and $\bar \psi$ are primaries with conformal dimension $1/2$, they transform as
\begin{equation}
    \psi(w) = \left( \frac{dw}{dz} \right)^{-1/2} \psi(z) , \qquad  \bar{\psi}(\bar{w}) = \left( \frac{d\bar{w}}{d\bar{z}} \right)^{-1/2} \bar{\psi}(\bar{z}).
\end{equation}
Combining this with Eq.~\eqref{eq:gluing-horizontal-defect-2} and noting that a point slightly above the defect on $\mathcal M_n$ (i.e. at $z =x + i 0^+$) is mapped to a point slightly to the left of the defect on the cylinder (i.e. at $w=x_j + i \tau + 0^-$), we find that the condition that $\psi$ and $\bar \psi$ must satisfy across the defect at the line $x_j=\frac{2\pi j}{n}$ on the cylinder $\mathcal{C}$ is 
\begin{equation}
\label{eq:gluing-cylinder}
  \Psi (w= x_j + i \tau + 0^- ) = R_{i \alpha_j} \Psi (w= x_j + i \tau + 0^+ ),
\end{equation}
where we define
\begin{equation}\label{eq:gluing-matrix}
\begin{aligned}
  R_{i\alpha_j} &=  
 \begin{pmatrix}
      \lr{\frac{dw}{dz}}^{-1/2} & 0\\
            0 &  \lr{\frac{d\bar w}{d\bar z}}^{-1/2}
       \end{pmatrix}
       \tilde R_{\alpha_j}  
       \begin{pmatrix}
            \lr{\frac{dw}{dz}}^{1/2} & 0\\
            0 &  \lr{\frac{d\bar{w}}{d\bar{z}}}^{1/2}
       \end{pmatrix} 
      =  \begin{pmatrix}
        \cos \alpha_j & i \sin \alpha_j \\
        i \sin \alpha_j & \cos \alpha_j
    \end{pmatrix}.
\end{aligned}
\end{equation}
Observe that, in the first equality, we can take out a factor $\left| \frac{dw}{dz} \right|^{-1/2}$ from the first matrix and a factor $\left| \frac{dw}{dz} \right|^{1/2}$ from the third one. Taking into account that $\frac{dw}{dz} / \left| \frac{dw}{dz} \right| = i$ for $z = x + i 0^+$, we find the second equality.

In summary, using the conformal transformation~\eqref{eq:conformal-map}, the partition function $Z_{\rm CFT}(\mathcal{M}_n^{\aalpha})$ of the massless Majorana fermion with $n$ marginal defect lines at the branch cuts of the surface $\mathcal{M}_n$ is equal to the partition function $Z_{\rm CFT}(\mathcal{C}^{\aalpha})$ of the theory on the cylinder $\mathcal{C}$ with $n$ equally-spaced defect lines along its longitudinal direction and described by the gluing conditions~\eqref{eq:gluing-cylinder}. If we impose conformal boundary conditions $\ket{a}$ and $\ket{b}$ at the extremes of the 
cylinder $\mathcal{C}$, the partition function $Z_{\rm CFT}(\mathcal{C}^{\aalpha})$ can be written as 
\begin{equation}\label{eq:partition_funct_cylinder}
 Z_{\rm CFT}(\mathcal{C}^{\boldsymbol{\alpha}})=\bra{a} e^{-W H} \ket{b},
\end{equation}
where $H$ is the Hamiltonian of the free Majorana 
fermion~\eqref{eq:majorana_ham} defined on a circle of length $2\pi$ 
with the fields $\psi$ and $\bar \psi$ satisfying the gluing 
conditions~\eqref{eq:gluing-matrix} at the points $x_j=\frac{2\pi j}{n}, \, j=1, \dots, n$ with strengths $\alpha=\alpha_1, \alpha_2, \dots, \alpha_n$ respectively. Alternatively, as we show in detail in 
Appendix~\ref{app:defect_Ham_formalism}, these conditions can be explicitly implemented in the 
Hamiltonian~\eqref{eq:majorana_ham} by including in it $n$ point defects of the form $\psi(x_j)\bar\psi(x_j)$,
\begin{equation}\label{eq:majorana_ham_defects}
	 H =\frac{1}{2 i}\int_0^{2\pi}  (\psi \partial_x\psi-\bar{\psi}\partial_x \bar{\psi})\d x  +
	 \sum_{j=1}^{n} i \mu_j \psi (x_j)\bar{\psi}(x_j), 
\end{equation}
where the the parameters $\mu_j$ are related to the the strength 
of the defects by $\mu_j/2=i\arctan(\alpha_j/2)$ see Appendix~\ref{app:defect_Ham_formalism} for a derivation.

For $W\gg 2\pi$, i.e. for large subsystem length $\ell$, the dominant term in the partition function~\eqref{eq:partition_funct_cylinder} is given by the the ground state 
energy $E(\boldsymbol{\alpha})$ of the Hamiltonian with defects~\eqref{eq:majorana_ham_defects},
\begin{equation}
 Z_{\rm CFT}(\mathcal{C}^{\boldsymbol{\alpha}})\propto e^{-W E(\boldsymbol{\alpha})}.
\end{equation}
The ground state energy should satisfy the usual CFT formula
\begin{equation}
 E(\aalpha)=\tilde{\Delta}_n(\aalpha)-\frac{c}{12},
\end{equation}
where $\tilde{\Delta}_n(\aalpha)$ takes into account the 
contribution of the defects and, consequently, it vanishes, 
$\tilde{\Delta}_n(\pmb 0)=0$, in their absence. It may be interpreted
as the scaling dimension of a $n$-defect insertion operator. Combining the two 
previous equations, and taking into account that $W=\frac{2}{n}\log(\ell/\epsilon)$, we arrive at Eq.~\eqref{eq:Delta_alpha}. 
Therefore, since $\tilde\Delta_n(\aalpha)=E(\boldsymbol{\alpha})-E(\boldsymbol{0})$, the problem of 
computing the scaling dimension $\tilde{\Delta}_n(\aalpha)$ boils down to determining the 
ground state energy of the Hamiltonian~\eqref{eq:majorana_ham_defects} with $n$ point defects. 
We will devote the rest of this section to calculating it.

However, before proceeding, it is important to note that the gluing condition~\eqref{eq:gluing-matrix} 
on the cylinder presents an issue: if $\alpha\in\mathbb{R}$, it does not respect the self-adjointness of the Majorana fields $\psi(w)$ 
and $\bar \psi(\bar w)$. The same problem arises in the Hamiltonian with defects~\eqref{eq:majorana_ham_defects}, which is not Hermitian for $\alpha\in\RR$.
To calculate $\tilde{\Delta}_n(\boldsymbol{\alpha})$, it is important that the Hamiltonian is Hermitian to ensure that its spectrum is real and, therefore, the energy of its ground state is well-defined. In order to cure this problem, we can analytically continue the defect strength $\alpha\to -i\lambda$ with $\lambda\in\mathbb{R}$.
This changes the $2\times 2$ gluing matrix \eqref{eq:gluing-matrix}
\begin{equation} \label{eq:gluing-matrix-lambda}
   R_{i \alpha} \rightarrow R_{\lambda}  =  \begin{pmatrix}
        \cosh \lambda & \sinh \lambda \\
        \sinh \lambda & \cosh \lambda
    \end{pmatrix}.
\end{equation}
Since all its entries are real, it is now compatible with the self-adjointness of the Majorana fields. This analytic continuation 
also makes the Hamiltonian with defects~\eqref{eq:majorana_ham_defects} Hermitian. In
the following, we carry out the calculation of the ground state energy assuming that the gluing 
matrix is~\eqref{eq:gluing-matrix-lambda} with $\lambda\in\mathbb{R}$. We will eventually take 
$\lambda\to i\alpha$ in the final result, which we check against exact numerical calculations in the XY spin chain.

\subsection{Ground state energy for a single defect (\texorpdfstring{$n=1$}{n=1})} 
\label{sec:FCS}

We start by solving the case of a single marginal defect. We take the spatial coordinate $x$ defined on the interval $x\in [-\pi, \pi]$ with the points $x=-\pi$ and $\pi$ identified and, for simplicity, we put the defect at $x=0$.
We impose the following boundary conditions for the fields $\psi(x)$, $\bar{\psi}(x)$ at $x=0$ and $x=\pi$: 
\begin{align}
    \Psi(0^-) &=  R_{\lambda} \Psi(0^+),  & \Psi(-\pi) = - \Psi(\pi).
\end{align}
The first one is the gluing condition for the defect, while the second one is the anti-periodic boundary condition. With these boundary conditions imposed on the fields, the Hamiltonian is 
\begin{equation}
    \label{eq:ham_n1}
    H=\frac{1}{2 i} \int_{-\pi}^0 \d x \, \Psi^\dagger D \Psi \, + \, \frac{1}{2 i} \int_0^{\pi} \d x \, \Psi^\dagger D \Psi, \quad \qquad  D  = \begin{pmatrix}
    \partial_x & 0 \\
        0 & - \partial_x
    \end{pmatrix} .
\end{equation}

\subsubsection{Diagonalization of the Hamiltonian}

The goal now is to diagonalize the Hamiltonian \eqref{eq:ham_n1}. To do this, we look for pairs of functions $(u(x),v(x))$ that satisfy the same gluing  and anti-periodic boundary conditions as $\Psi(x)$,
\begin{equation}\label{eq:bc_u_v}
    \begin{pmatrix}
        u(0^-) \\ v(0^-) 
    \end{pmatrix} =  R_{\lambda} 
    \begin{pmatrix}
    u(0^+) \\ v(0^+) 
    \end{pmatrix},
    \quad 
     \begin{pmatrix}
        u(-\pi) \\ v(-\pi) 
    \end{pmatrix} = -
    \begin{pmatrix}
    u(\pi) \\ v(\pi) 
    \end{pmatrix},
\end{equation}
and are eigenstates of the differential operator $\frac{1}{i} D$. These are piecewise plane waves, 
\begin{align}
u_k(x) &=
    \begin{cases}
        A_0 e^{ikx}, & x \in (-\pi,0), \\
        A_1 e^{ikx}, & x \in (0, \pi),
    \end{cases}
    &
    v_k(x) &=
    \begin{cases}
        B_0 e^{-ikx}, & x \in (-\pi,0), \\
        B_1 e^{-ikx}, & x \in (0, \pi) .
    \end{cases}
\end{align}
If they satisfy the boundary conditions, then such wavefunctions are automatically eigenfunctions of $\frac{1}{i}D$ with eigenvalue $k$.

The conditions~\eqref{eq:bc_u_v} impose the following constraints on the amplitudes:
\begin{align}
    \begin{pmatrix}
        A_0 \\ B_0
    \end{pmatrix}
    &=
     R_\lambda 
    \begin{pmatrix}
        A_1 \\ B_1
    \end{pmatrix},
    &
    \begin{pmatrix}
        A_0 e^{-ik\pi} \\ B_0 e^{ik\pi}
    \end{pmatrix}
    &=
    -\begin{pmatrix}
        A_1 e^{ik\pi} \\ B_1 e^{-ik\pi}
    \end{pmatrix} .
\end{align}
This linear system of equations admits a non-zero solution if and only if
\begin{equation}
    \label{eq:constraint_n1}
    \det \slr{ \mathbb{I} +  R_{\lambda} \begin{pmatrix}
        e^{-ik2\pi} & 0 \\
        0 & e^{ik2\pi}
    \end{pmatrix}  }
    = 0 .
\end{equation}
Let us introduce the polynomial 
\begin{equation}
P_\lambda (z)  =  z \, \det \slr{  \mathbb{I} + R_{\lambda} \begin{pmatrix}
        1/z & 0 \\
        0 & z
    \end{pmatrix} }
\end{equation}    
of degree $2$. Eq.~(\ref{eq:constraint_n1}) is equivalent to the polynomial equation 
\begin{equation}
    \label{eq:constraint_n1_pol}
    P_\lambda (z) = 0
\end{equation}
for the variable $z = e^{i 2\pi k}$.
From the explicit form of $R_\lambda$, we find that $P_\lambda(z)$ is
\begin{equation}
    P_\lambda (z) = {\rm const.} \times (z - e^{i \theta}) (z - e^{-i \theta}) ,
\end{equation}
with
\begin{equation}
   \theta = 2\arctan \lr{\tanh \frac{\lambda}{2}} + \pi  \; \in  [0, 2\pi) . 
\end{equation}
Then the full set of solutions $k$ of Eq.~(\ref{eq:constraint_n1}) is
\begin{equation}\label{eq:solutions_one_defect}
    k \in \mathcal{S}_\lambda =  \left( \mathbb{Z} + \frac{\theta}{2\pi}   \right) \cup \left( \mathbb{Z}  - \frac{\theta}{2\pi}   \right) .
\end{equation}
For each solution $k \in \mathcal{S}_\lambda$, the pair $(u_k,v_k)$ can be used to construct a Bogoliubov mode for the Hamiltonian \eqref{eq:ham_n1}, taking the scalar product with the two-component field $(\psi, \bar \psi)$
\begin{equation}
    \label{eq:bogo_etak1}
    \eta_k = \int_{-\pi}^{\pi} \d x \, [u^*_k(x) \psi(x) + v^*_k(x) \bar{\psi}(x)], 
\end{equation}
which automatically satisfies $[H, \eta_k] \, = \, k \, \eta_k $.
Then, using the orthonormality of the set of functions $(u_k(x), v_k(x))$ we get
\begin{equation}\label{eq:ham_one_defect_diag_dagger}
    H = \frac{1}{2} \sum_{k\in \mathcal S_\lambda} k \eta^\dagger_k \eta_k,
\end{equation}
where the sum in $k$ runs over all the solutions in Eq.~\eqref{eq:solutions_one_defect} and the modes satisfy the anticommutation relations $\{ \eta^\dagger_k, \eta_q \} = \delta_{k,q}$.

Notice that, taking the complex conjugate of the eigenvalue equation for $\frac{1}{i}D$, we get that $(u_k^*, v_k^*)$ is an eigenvector with eigenvalue $-k$,
\begin{equation}
    D \begin{pmatrix}
        u_k \\ v_k
    \end{pmatrix}
    = ik 
    \begin{pmatrix}
        u_k \\ v_k
    \end{pmatrix}
    \implies
    D \begin{pmatrix}
        u^*_k \\ v^*_k
    \end{pmatrix}
    = -ik 
    \begin{pmatrix}
        u^*_k \\ v^*_k
    \end{pmatrix}.
\end{equation}
Thus we can set $u_{-k}(x) = u_k^*(x)$ and $v_{-k}(x) = v_k^*(x)$. This implies that $\eta_{-k} = \eta^\dagger_k$, and Eq.~\eqref{eq:ham_one_defect_diag_dagger} can be rewritten in the form
\begin{equation}
    H = \frac{1}{2} \sum_{k \in \mathcal{S}_\lambda} k \, \eta_k \eta_{-k}.
\end{equation}
Alternatively, we can express it as a sum restricted to the set of positive solutions $k$, $\mathcal{S}^+_\lambda = \{ k \in \mathcal{S}_\lambda \mid k >0  \}$,
\begin{equation}
    \label{eq:H_Slambdaplus}
    H =   \sum_{k \in \mathcal{S}^+_\lambda} k \, ( \eta_k \eta_{-k}  - 1/2  ).
\end{equation}

\subsubsection{Ground state energy}

From Eq.~\eqref{eq:H_Slambdaplus}, it is clear that the ground state of the single-defect Hamiltonian~\eqref{eq:ham_n1} is the state annihilated by all the modes $\eta_{-k}$ for $k \in \mathcal{S}^+_\lambda$. The ground state energy is
\begin{equation}
    E(\lambda) =  - \frac{1}{2} \sum_{k \in \mathcal{S}^+} k = -\frac{1}{2} \slr{ \sum_{m =0 }^\infty \left(m + \frac{\theta}{2\pi}\right)  + \sum_{m=1}^\infty \left(m - \frac{\theta}{2\pi}\right) }.
\end{equation}
These infinite sums can be evaluated by zeta-regularization. Taking into account that
\begin{equation}\label{eq:zeta_regularization}
  \lim_{s \rightarrow -1} \sum_{m =0 }^\infty (m + a)^{-s}=\zeta(-1, a),
\end{equation}
where $\zeta(s, a)=\sum_{m=0}^\infty(m+a)^{-s}$ is the Hurtwitz zeta function, the ground state energy can be written as
\begin{equation}
   E(\lambda) =  -\frac{1}{2}   \left[  \zeta\left(-1,  \frac{\theta}{2\pi}\right) + \zeta\left(-1,  1- \frac{\theta}{2\pi}\right) \right].
\end{equation}
Using the identity $\zeta(-1, a) =  -\frac{1}{12} + \frac{a}{2} - \frac{a^2}{2} = \zeta(-1,1-a)$, we arrive at
\begin{equation}
     E(\lambda) =   \frac{1}{2}  \lr{ \frac{1}{\pi} {\arctan} \left(  \tanh \frac{\lambda}{2} \right) }^2 -\frac{1}{24}  .
\end{equation}
Identifying this expression with the standard formula for the ground state energy in a CFT,
\begin{equation}
    E(\lambda) =  \Delta_1(\lambda) - \frac{c}{12} ,
\end{equation}
with $c=1/2$ for the massless Majorana fermion, we find that the scaling dimension associated with the insertion of a single marginal defect of strength $\lambda$ is
\begin{equation}\label{eq:scaling-dim-one-defect}
    \Delta_1(\lambda)  = \frac{1}{2\pi^2}  \arctan^2 \lr{  \tanh \frac{\lambda}{2} } . 
\end{equation}

\subsubsection{Connection with previous works}
The scaling dimension associated with a single defect was computed in Ref.~\cite{hps-89} applying lattice methods in the quantum Ising chain and in Ref.~\cite{oa-97} using a boundary CFT approach. In the latter, the Ising CFT with a defect is folded along the defect, obtaining a $\ZZ_2$ orbifold of the compact boson in which the defect is encoded in the boundary condition. The relation between such bosonic boundary condition and our gluing parameter $\lambda$ can be found in Ref.~\cite{bb-15}. 

In the case $n=1$, the charged moments~\eqref{eq:def-Z(g)} 
specialize to $\mathcal{Z}_1(\alpha)={\rm Tr}(\rho_A e^{i\alpha Q_A})$. This is the full counting statistics, i.e. the cumulant 
generating function, of the charge $Q_A$ in the subsystem
$A$. In our setup, it corresponds to the expectation value of a single defect line on the single replica surface $\mathcal{M}$.
In the ground state of the critical XY spin chain, this quantity was  calculated in Refs.~\cite{cd-2007,gec-18} employing lattice
methods, see also~\cite{Stephan:2013logarithmic, ia-13, arv-21}, and obtaining that 
$\mathcal{Z}_1(\alpha)=e^{-t(\alpha)\ell}\ell^{-2\tilde\Delta_1(\alpha)}$, with $t(\alpha)$ 
given by Eq.~\eqref{eq:t_n_xy}, and the exponent $\tilde\Delta_1(\alpha)=\Delta_1(-i\alpha)$ that we have found in Eq.~\eqref{eq:scaling-dim-one-defect} using CFT.

Notice that the case $\lambda=i\pi$ corresponds to the $\mathbb{Z}_2$ spin-flip symmetry of the spin chain
and the associated defect is topological --- the $\varepsilon$ Verlinde line~\cite{pz-00, ffrs-04, yin-19}. When this line is open, as in our case,
 two disorder operators $\mu(z)$ are inserted at its end-points~\cite{ffrs-04}. Therefore, by the Kramers-Wannier 
duality, $\mathcal{Z}_1(\pi)$ should be proportional to the two-point function of the spin fields $\sigma(z)$ at its end-points, 
$\mathcal{Z}_1(\pi)\propto \langle\sigma(0)\sigma(\ell)\rangle=\ell^{-2\Delta_\sigma}$, which have 
scaling dimension $\Delta_\sigma=1/8$, that is precisely $\Delta_1(i\pi)$.

\subsection{Ground state energy for \texorpdfstring{$n$}{n} equally-spaced defects}
\label{sec:gs-energy-n-defects}

We now extend the calculation of the previous section to the case of multiple defects. In this section we take the spatial coordinate $x$ in the interval $[0,2\pi]$, and we put the defects at positions $x_j=\frac{2\pi j}{n}$ with $j=1, \dots, n$. We also define $x_0 = 0$. The Hamiltonian is
\begin{equation}
    \label{eq:ham_n}
    H= \sum_{j=1}^{n} \frac{1}{2 i} \int_{x_{j-1}}^{x_j} \d x \, \Psi^\dagger D \Psi ,  \quad \qquad  D  = \begin{pmatrix}
    \partial_x & 0 \\
        0 & - \partial_x
    \end{pmatrix} ,
\end{equation}
with the following gluing conditions corresponding to $n$ equally-spaced defects of strengths $\lambda_1$, $\lambda_2$, \dots, $\lambda_n$,
\begin{align} \label{eq:gluing-cond-ndef}
    \Psi( x_j + 0^-) &=  R_{\lambda_j} \Psi( x_j + 0^+)  ,
\end{align}
where the matrix $R_\lambda$ was defined in Eq.~\eqref{eq:gluing-matrix-lambda}, and the anti-periodic boundary condition $\Psi(0) = - \Psi(2\pi)$.

\subsubsection{Diagonalization of the Hamiltonian}

To diagonalize that Hamiltonian, we proceed as in the $n=1$ case. We look for pairs of functions $(u_k(x),v_k(x))$ that satisfy the same gluing conditions as $\Psi(x)$ and are eigenstates of $\frac{1}{i}D$. We look for solutions in the form of piecewise plane waves,
\begin{align}
    u_k(x) &=
    \begin{cases}
        A_0 e^{ikx}, & x \in (0,x_1), \\
        A_1 e^{ikx}, & x \in (x_1,x_2), \\
        \vdots \\
        A_n e^{ikx}, & x \in (x_{n-1}, x_n),
    \end{cases}
    &
    v_k(x) &=
    \begin{cases}
        B_0 e^{-ikx}, & x \in (0,x_1), \\
        B_1 e^{-ikx}, & x \in (x_1,x_2), \\
        \vdots \\
        B_n e^{-ikx}, & x \in (x_{n-1}, x_n) .
    \end{cases}
\end{align}
The gluing conditions~\eqref{eq:gluing-cond-ndef} imply the following relations between consecutive amplitudes
\begin{equation}\label{eq:gluing-jwithj-1}
    \begin{pmatrix}
        A_{j-1} \\ B_{j-1}
    \end{pmatrix}
    = 
    \begin{pmatrix}
        e^{-ikx_j} & 0 \\
        0 & e^{ikx_j}
    \end{pmatrix}
    R_{\lambda_j} 
    \begin{pmatrix}
        e^{ikx_j} & 0 \\
        0 & e^{-ikx_j}
    \end{pmatrix}
    \begin{pmatrix}
        A_{j} \\ B_{j}
    \end{pmatrix},
\end{equation}
while the anti-periodicity condition implies
\begin{equation}
    \begin{pmatrix}
        A_{0} \\ B_{0}
    \end{pmatrix}
    = 
    - 
    \begin{pmatrix}
        e^{ik2\pi} & 0 \\
        0 & e^{-ik2\pi}
    \end{pmatrix}
    \begin{pmatrix}
        A_{n} \\ B_{n}
    \end{pmatrix} .
\end{equation}
This system of equations admits a non-zero solution if and only if $k$ is such that

\begin{equation}
\label{eq:quantization-condition}
    \det \lr{ \mathbb{I} + R_{\lambda_1} \begin{pmatrix}
        e^{-i\frac{k2\pi}{n}} & 0 \\
        0 & e^{i\frac{k2\pi}{n}}
    \end{pmatrix}  R_{\lambda_2} \begin{pmatrix}
        e^{-i\frac{k2\pi}{n}} & 0 \\
        0 & e^{i\frac{k2\pi}{n}}
    \end{pmatrix} \dots R_{\lambda_{n}} \begin{pmatrix}
        e^{-i\frac{k2\pi}{n}} & 0 \\
        0 & e^{i\frac{k2\pi}{n}}
    \end{pmatrix}  } = 0.
\end{equation}
It is convenient to define the polynomial
\begin{equation}\label{eq:P_lambda_n_defects}
    P_{\llambda} (z) = z^{n}  \det \lr{ \mathbb{I} + R_{\lambda_1} \begin{pmatrix}
        1/z & 0 \\
        0 & z
    \end{pmatrix}  R_{\lambda_2} \begin{pmatrix}
        1/z & 0 \\
        0 & -z
    \end{pmatrix} \dots R_{\lambda_{n}} \begin{pmatrix}
        1/z & 0 \\
        0 & z
    \end{pmatrix}  }, 
\end{equation}
of degree $2n$. This polynomial is palindromic, i.e. it satisfies $P_{\llambda}(z) = z^{2n} P_{\llambda}(1/z)$, and it has real coefficients. Let us call $z_j$, $j=1,\dots, 2n$, the roots of that polynomial. When $\llambda \in \RR^n$, the roots lie on the unit circle $|z_j|=1$. This corresponds to having real solutions for $k$ in Eq.~\eqref{eq:quantization-condition}. Therefore, in this case, we can write $z_j = e^{i \theta_j}$ with $\theta_j \in [0,2\pi)$. Later, when we will analytically continue our result, the relation between $z_j$ and $\theta_j$ will just be $\theta_j = -i \Log z_j$, without the property $\theta_j \in \RR$. 

The polynomial~\eqref{eq:P_lambda_n_defects} can be rewritten in terms of its roots
\begin{equation}
    P_{\llambda} (z)  = \mathrm{const.} \times \prod_{j=1}^{2n} (z-e^{i \theta_j({\llambda})}), \qquad \theta_j({\llambda}) \in [0, 2\pi ),
\end{equation}
and each root determines a family of solutions to the quantization condition \eqref{eq:quantization-condition} for $k$ via $z_j = e^{ik2\pi/n}$. The set of all such solutions is then
\begin{eqnarray}
    \mathcal{S}_{\llambda} = \bigcup_{j=1}^{2n}n \left(\mathbb{Z}  + \frac{\theta_j({\llambda})}{2\pi}\right)  .
\end{eqnarray}
Each $k\in \mathcal{S}_{\llambda}$ defines a Bogoliubov mode
\begin{equation}
    \label{eq:bogo_etakn}
    \eta_k = \int_{0}^{2\pi} \d x \, [u^*_k(x) \psi(x) + v^*_k(x) \bar{\psi}(x)], 
\end{equation}
which automatically satisfies $[H, \eta_k] \, = \, k \, \eta_k $ along with $(\eta_k)^\dagger = \eta_{-k}$, and the canonical anticommutation relations $\{ \eta^\dagger_k, \eta_q \} = \delta_{k,q}$. Then the $n$-defect Hamiltonian~\eqref{eq:ham_n} is diagonal in terms of them,
\begin{equation}
    H = \frac{1}{2} \sum_{k\in \mathcal S_\lambda} k \eta^\dagger_k \eta_k,
\end{equation}
where the sum runs over all the solutions $k$ of Eq.~\eqref{eq:quantization-condition}. Alternatively, one can write it as
\begin{equation}\label{eq:ham_n_diag}
    H =  \sum_{k \in \mathcal{S}^+_{\llambda} } k (\eta_k \eta_{-k} - 1/2),
\end{equation}
where the sum is now restricted to the set of positive solutions,
$\mathcal{S}^+_{\llambda} = \{ k \in \mathcal{S}_{\llambda} | k >0 \}$.
This expression is particularly convenient to compute the ground state energy.

\subsubsection{Ground state energy}

According to Eq.~\eqref{eq:ham_n_diag}, the ground state of the $n$-defect Hamiltonian~\eqref{eq:ham_n} corresponds to the configuration with all the positive modes $k$ occupied. Its energy is
\begin{equation}
    E({\llambda}) = -\frac{1}{2} \sum_{{k \in \mathcal{S}^+_{\llambda}}} k=-\frac{1}{2} \sum_{j=1}^{2n} \sum_{m = 0}^\infty n \left(m +  \frac{\theta_j}{2\pi}\right).
\end{equation}
As in the $n=1$ case above, these divergent series can be evaluated by zeta-regularization, using Eq.~\eqref{eq:zeta_regularization},
\begin{equation}
 E(\llambda) = -\frac{n}{2}  \sum_{j=1}^{2n}  \zeta\left(-1,  \frac{\theta_j}{2\pi}\right).
\end{equation}
If we apply the identity $\zeta(-1, a)  =  \frac{1}{24} - \frac{(a- 1/2)^2}{2}$ for the Hurwitz zeta function, then we obtain
\begin{equation}
 E(\llambda) =  \frac{n}{2}  \sum_{j=1}^{2n} \left(  \frac{1}{2} \left( \frac{\theta_j}{2\pi} - \frac{1}{2} \right)^2 - \frac{1}{24} \right).
\end{equation}
This expression should be identified with the usual CFT formula for the ground state energy, $E =\Delta_n({\llambda}) - c/12 $, with $c=1/2$. Therefore, we find that for $n$ equally-spaced defects the scaling dimension $\Delta_n(\llambda)$ is
\begin{equation}
    \label{eq:Delta_lambda_n}
    \Delta_n(\llambda) = -\frac{n^2-1}{24} + \frac{n}{4}  \sum_{j=1}^{2n}  \left( \frac{\theta_j}{2\pi} - \frac{1}{2} \right)^2   .
\end{equation}
This is the second main result of this paper section, which we will use to derive 
the R\'enyi entanglement asymmetry of the critical XY spin chain in 
the next section, where we also report the explicit expression of 
$\Delta_n(\llambda)$ for $n=2$ and $3$. As a first check 
of Eq.~\eqref{eq:Delta_lambda_n}, note that, when 
$\lambda_1=\lambda_2=\dots=\lambda_n=0$, we must obtain 
$\Delta_n(\pmb 0)=0$. In fact, in that case, the 
polynomial~\eqref{eq:P_lambda_n_defects} is
$P_{\boldsymbol{\lambda}}(z)=(z^n+1)^2$ and its roots are $z_j=e^{i 2\pi \frac{j-1/2}{n}}$, 
$j=1,\dots, n$, all with multiplicity $2$. 
Therefore, we have $\theta_j =  \theta_{j+n}=2\pi (j-1/2)/n$ for $1 \leq j \leq n $. Inserting 
these roots in Eq.~\eqref{eq:Delta_lambda_n} and performing the sum, we find $\Delta_n(\pmb 0)=0$ as it should be.

\subsection{Summary}
\label{sec:summary-n-defects}

For the convenience of the reader, let us briefly summarize the main result of this section. It is the the second important result of this paper. It gives the exact scaling dimension $\Delta_n(\boldsymbol{\lambda})$ associated with the insertion of $n$-equally-spaced marginal defects in the massless Majorana fermion on a circle, with strengths $\lambda_1, \dots, \lambda_n$. The result is given by
Eq.~\eqref{eq:Delta_lambda_n}, which can also be rewritten in the form
\begin{equation}\label{eq:scaling-dimension-result-summary}
    \Delta_n (\llambda) = -\frac{n^2-1}{24} + \frac{n}{4}  \sum_{j=1}^{2n}  \lr{ \frac{\Log(-z_j)}{2\pi i} }^2,
\end{equation}
where $\Log(.)$ is the principal value of the logarithm, whose imaginary part takes values in $(-\pi,\pi]$ and its branch cut is taken along the negative real axis, and the $z_j$'s ($j = 1, \dots, 2n$) are the $2n$ roots of the following polynomial of degree $2n$:
\begin{equation}\label{eq:polynomial-n-defects2}
    P_{\llambda} (z) = z^{n}  \det \lr{ \mathbb{I} + R_{\lambda_1}    
    \begin{pmatrix}
        1/z & 0 \\
        0 & z
    \end{pmatrix}  R_{\lambda_2} \begin{pmatrix}
        1/z & 0 \\
        0 & z
    \end{pmatrix} \dots R_{\lambda_{n}} \begin{pmatrix}
        1/z & 0 \\
        0 & z
    \end{pmatrix}  }, 
\end{equation}
with 
\begin{equation}
R_\lambda = \begin{pmatrix}
        \cosh \lambda & \sinh \lambda \\
        \sinh \lambda & \cosh \lambda
\end{pmatrix}.
\end{equation}
We have derived this result for defect strengths $\lambda_j \in \mathbb{R}$, but it can be analytically continued to $\lambda_j \in \mathbb{C}$. In particular, in what follows, we will take $\lambda_j \rightarrow i \alpha_j$  to derive the R\'enyi entanglement asymmetry in the critical XY spin chain.

\section{R\'enyi entanglement asymmetry in the critical XY spin
chain}\label{sec:asymm_XY}

In this section, we derive the asymptotic behavior of the R\'enyi 
entanglement asymmetry in the ground state of the critical XY spin 
chain using the results obtained above. At the critical lines 
$\gamma\neq 0$, $|h|=1$, the charged moments $\mathcal{Z}_n(\aalpha)$ behave as in Eq.~\eqref{eq:Zng-critical} for large subsystem 
length $\ell$. While the string tension $T_n(\aalpha)$ is given 
by Eq.~\eqref{eq:t_n_xy}, we have found in Sec.~\ref{sec:gs-energy-n-defects} that the scaling dimension $\tilde{\Delta}_n(\aalpha)$ 
can be obtained, upon the analytic continuation $\boldsymbol{\lambda}=i\boldsymbol{\alpha}$, from 
Eq.~\eqref{eq:scaling-dimension-result-summary}, which further 
requires to determine the roots of the polynomial~\eqref{eq:polynomial-n-defects2}. 
Unfortunately, we are not able to find a general expression for these roots. Here we first consider the cases $n=2$ and $3$, and we check our analytic prediction for $\mathcal{Z}_n(\boldsymbol{\alpha})$ against exact numerical results in the ground state of the XY spin chain. We then derive by applying the saddle point approximation discussed in Sec.~\ref{sec:generic_asymp_beh} the asymptotic behavior of the R\'enyi entanglement asymmetry for any integer index $n$, and by analytically continuing it, the replica limit $n\to 1$.

\subsection{Numerical checks}

\subsubsection{\texorpdfstring{$n=2$}{n=2} charged moments}
\label{sec:n=2}

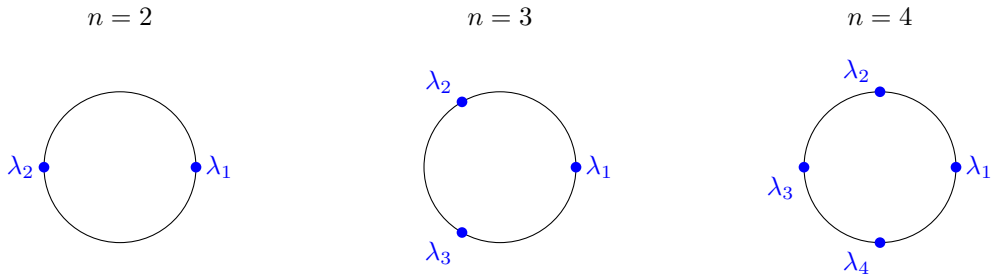
\begin{figure}
\begin{center}
\begin{tikzpicture}
  \coordinate (O) at (0,0); 
  \coordinate (A) at (0:1cm); 
  \coordinate (B) at (180:1cm); 

  \draw (O) circle[radius=1cm];
  \fill[blue] (A) circle[radius=2pt] node[right] {$\lambda_1$};
  \fill[blue] (B) circle[radius=2pt] node[left] {$\lambda_2$};
  \node at (0,2) {$n=2$};

  \begin{scope}[xshift=5cm]
  \coordinate (O) at (0,0); 
  \draw (O) circle[radius=1cm];
  \coordinate (A) at (0:1cm);  
  \coordinate (B) at (120:1cm); 
  \coordinate (C) at (240:1cm); 
  \fill[blue] (A) circle[radius=2pt] node[right] {$\lambda_1$};
  \fill[blue] (B) circle[radius=2pt] node[above left] {$\lambda_2$};
  \fill[blue] (C) circle[radius=2pt] node[below left] {$\lambda_3$};
  \node at (0,2) {$n=3$};
  \end{scope}

  \begin{scope}[xshift=10cm]
  \coordinate (O) at (0,0); 
  \draw (O) circle[radius=1cm];
  \coordinate (A) at (0:1cm);  
  \coordinate (B) at (90:1cm); 
  \coordinate (C) at (180:1cm); 
  \coordinate (D) at (270:1cm); 
  \fill[blue] (A) circle[radius=2pt] node[right] {$\lambda_1$};
  \fill[blue] (B) circle[radius=2pt] node[above left] {$\lambda_2$};
  \fill[blue] (C) circle[radius=2pt] node[below left] {$\lambda_3$};
  \fill[blue] (D) circle[radius=2pt] node[below left] {$\lambda_4$};
  \node at (0,2) {$n=4$};
  \end{scope}
\end{tikzpicture}
\end{center}
\caption{Disposition of the point defects on the circle in the calculation of the charged moments $\mathcal{Z}_n(\aalpha)$ for $n=2$, $3$ and $4$. If the circle has length $2\pi$, then the defect of strength $\lambda_j$ is located at the point $x_j=\frac{2\pi j}{n}$.}
\label{fig:defects_n_2_3_4}
\end{figure}

In the case of two defects located at the points indicated in the left panel of Fig.~\ref{fig:defects_n_2_3_4}, the polynomial of \cref{eq:P_lambda_n_defects} reads
\begin{equation}
     P_{\llambda}(z) = \cosh \lambda_1 \cosh\lambda_2 (1+z^4) +2(1+\sinh\lambda_1 \sinh\lambda_2) z^2.
\end{equation}
It is a bit cumbersome to write the roots explicitly, but using them in Eq.~(\ref{eq:scaling-dimension-result-summary}) we arrive at the formula for the scaling dimension associated with the insertion of two defects
\begin{equation} \label{eq:scaling-dimension-2-two}
    \Delta_2(\lambda_1,\lambda_2) = \frac{1}{2\pi^2} \slr{ \arctan \lr{ \tanh \frac{\lambda_1}{2} } - \arctan \lr{ \tanh \frac{\lambda_2}{2} } }^2. 
\end{equation}
To keep formulas compact, here we write the roots only in the special case $\lambda_2 = -\lambda_1 = \lambda$, which is the case that we use below in our analysis of the asymmetry. In that case the four roots are:
\begin{equation}
    z = \pm \frac{i \pm \sinh \lambda}{\cosh \lambda},
\end{equation}
and, taking $\theta=-i\log z$, their arguments are
\begin{align}
    \theta_{1} &= \pi-\theta_2 =2\arctan \tanh(\lambda/2) + \frac{\pi}{2}, \\
    \theta_{3} &=3\pi-\theta_4= 2\arctan \tanh(\lambda/2) + \frac{3\pi}{2}.
\end{align}
Plugging them in Eq.~\eqref{eq:scaling-dimension-result-summary}, we obtain
\begin{equation}
    \Delta_2(\lambda,-\lambda) = \frac{2}{\pi^2} \arctan^2 \lr{ \tanh \frac{\lambda}{2} }
\end{equation}
and, taking the analytic continuation $\lambda = i \alpha$,
\begin{equation}\label{eq:scaling-dimension-2}
    \tilde\Delta_2(\alpha,-\alpha) = -\frac{2}{\pi^2} \arctanh^2 \lr{ \tan \frac{\alpha}{2} }.
\end{equation}

Note that, in Eq.~\eqref{eq:scaling-dimension-2}, $\tilde{\Delta}_2
(\alpha,-\alpha)$ is only well-defined in the interval $\alpha\in(-\pi/2, 
\pi/2)$ since the domain of definition of $\arctanh(x)$ is $x\in (-1, 1)$. On the other hand, the ground state of the critical XY spin 
chain is invariant under the subgroup $\ZZ_2 \subset U(1)$ of spin 
flips, which implies that the charged moments $\mathcal{Z}_2(\aalpha)$ 
are periodic $\mathcal{Z}_2(\aalpha+\boldsymbol\pi)=\mathcal{Z}_2(\aalpha)$. 
Therefore, Eq.~\eqref{eq:scaling-dimension-2} must be extended 
outside the interval $\alpha\in (-\pi/2, \pi/2)$ such that this 
periodicity is satisfied,
\begin{equation}\label{eq:scaling-dimension-2-periodic}
 \tilde{\Delta}_2(\alpha,-\alpha)=
 \begin{cases}
  -\frac{2}{\pi^2} \arctanh^2\left(\tanh\frac{\alpha}{2}\right), & \alpha\in(-\pi/2, \pi/2), \\
  -\frac{2}{\pi^2} \arctanh^2\left(\tanh\frac{(\pi-\alpha)}{2}\right), & \alpha\in(-\pi, -\pi/2)\cup(\pi/2, \pi).
 \end{cases}
\end{equation}
In the left panel of Fig.~\ref{fig:charged-moment-n2}, we numerically check this result. As we explain in Appendix~\ref{app:numerics}, the charged moments $\mathcal{Z}_n(\aalpha)$ can 
be exactly calculated numerically in the ground state of the critical 
XY spin chain with Eq.~\eqref{eq:numerics}. Using this expression together with Eq.~\eqref{eq:t_n_xy}, we compute $\log(\mathcal{Z}_2(\aalpha)e^{T_2(\aalpha)\ell})$ with $\aalpha= (\alpha,-\alpha)$ for a fixed $\alpha$ and $\ell=50, 60,\dots, 100$ and we fit the curve
$-\tilde{\Delta}_2(\aalpha)\log\ell+{\rm const.}$ to this set of points. 
In the plot on the left side of Fig.~\ref{fig:charged-moment-n2}, the symbols correspond to the values of $\tilde{\Delta}_2(\aalpha)$ obtained in the fit for different angles $\alpha$ and couplings ($h=1, \gamma)$, while the solid curve is the prediction of Eq.~\eqref{eq:scaling-dimension-2-periodic}. We obtain a very good agreement between them.

The divergence of $\tilde\Delta_2(\aalpha)$ in $\aalpha = (\pm \pi/2, \mp \pi/2)$ does not mean that the charged moment is divergent itself but that the charged moment has a different scaling in $\ell$. We numerically observe that in this case the scaling is $\log \mathcal Z_2(\aalpha) = - T(\aalpha) \ell + O\lr{(\log \ell)^2}$. In general, we observe this anomalous scaling with a $(\log \ell)^2$ term in the charged moment $\mathcal Z_n(\aalpha)$ for every $n$ when at least one $\alpha_j$ is equal to $\pi/2$. Being these points a measure zero set in the integral for the asymmetry, the analysis performed in \cref{sec:generic_asymp_beh} does not change.

\begin{figure}
\centering
\includegraphics[width=0.49\textwidth]{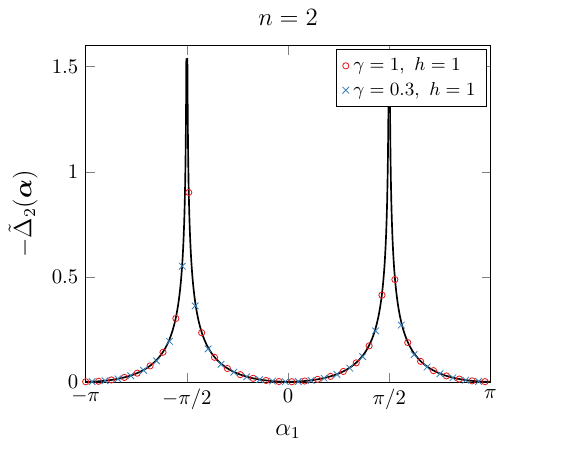}
\includegraphics[width=0.49\textwidth]{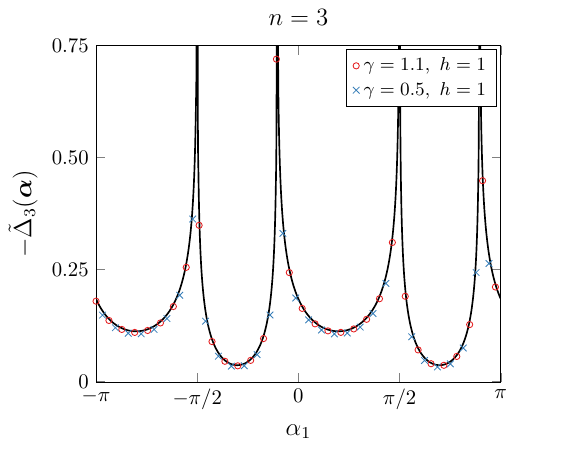}
\caption{Scaling dimension $\tilde{\Delta}_n(\aalpha)$ for two (left panel) and three (right panel) defects, which appears in the asymptotic behavior of the charged moments $\mathcal{Z}_n(\aalpha)$. For $n=2$, we take $\alpha_2=-\alpha_1$ and we vary $\alpha_1$. For $n=3$, we set $\alpha_1+\alpha_2+\alpha_3=0$ and change $\alpha_1$ with $\alpha_2=1.9$. The symbols have been obtained numerically as detailed in the main text for the ground state of the XY spin chain along the critical line $\gamma>0$ and $h=1$. The curves are the CFT prediction~\eqref{eq:scaling-dimension-result-summary}, that for $n=2$ simplifies to~\eqref{eq:scaling-dimension-2-periodic}.}
\label{fig:charged-moment-n2}
\end{figure}

\subsubsection{\texorpdfstring{$n=3$}{n=3} charged moments}

For three defects at the positions of the middle panel of Fig.~\ref{fig:defects_n_2_3_4} on a circle, the polynomial \eqref{eq:P_lambda_n_defects} is
\begin{equation}
    P_{\llambda}(z) = C_{\boldsymbol{\lambda}} + S_{\boldsymbol{\lambda}} z^2 + 2 z^3 + S_{\boldsymbol{\lambda}} z^4 + C_{\boldsymbol{\lambda}} z^6,
\end{equation}
with 
\begin{equation}
    C_{\boldsymbol{\lambda}} = \cosh \lambda_1 \cosh \lambda_2 \cosh \lambda_3,\quad
    S_{\boldsymbol{\lambda}} = \sinh \lambda_1 \sinh \lambda_2 \cosh \lambda_3 + \text{cycl. perm.}
\end{equation}
To compute the coefficient $\tilde{\Delta}_3(\aalpha)$ that enters
in the asymptotic behavior of the charged moment $\mathcal{Z}_3(\aalpha)$, we have to impose $\lambda_1+\lambda_2+\lambda_3=0$ due to the Dirac delta in \cref{eq:renyi_ent_asymm_charged_mom}. In that case, $S_{\boldsymbol{\lambda}}=1-C_{\boldsymbol{\lambda}}$ and the polynomial has two equal roots $z_1=z_2=-1$. The other four roots are 
\begin{equation}\label{eq:roots_n_3}
\begin{aligned}
    z_{3} &= z_4^*= \frac{ \sqrt{C}_{\boldsymbol{\lambda}} + \sqrt{C_{\boldsymbol{\lambda}}-1} \pm i \sqrt{1 + 2C_{\boldsymbol{\lambda}} - 2 \sqrt{C_{\boldsymbol{\lambda}}(C_{\boldsymbol{\lambda}}-1)}} }{2\sqrt C_{\boldsymbol{\lambda}}}, \\
    z_{5} &= z_6^*= \frac{ \sqrt{C}_{\boldsymbol{\lambda}} - \sqrt{C_{\boldsymbol{\lambda}}-1} \pm i \sqrt{1 + 2C_{\boldsymbol{\lambda}} + 2 \sqrt{C_{\boldsymbol{\lambda}}(C_{\boldsymbol{\lambda}}-1)}} }{2\sqrt C_{\boldsymbol{\lambda}}}.
\end{aligned}
\end{equation}
Plugging them in Eq.~\eqref{eq:scaling-dimension-result-summary} and performing the analytic continuation
$\llambda=i\pmb\alpha$, we obtain the analytic expression for $\tilde{\Delta}_3(\aalpha)$. We numerically check it in the right panel of Fig.~\ref{fig:charged-moment-n2} as
we have done for the case $n=2$. We can calculate the exact value of the charged moment $\mathcal{Z}_3(\aalpha)$ in the critical XY spin chain employing Eq.~\eqref{eq:numerics} in the appendix. Combining it with Eq.~\eqref{eq:t_n_xy}, we compute $\log(\mathcal{Z}_3(\aalpha)e^{\ell T_3(\aalpha)})$ for a given $\aalpha=(\alpha_1, \alpha_2, -\alpha_1-\alpha_2)$ and $\ell=50, 60, \dots, 100$. With the resulting set of points, we fit the function $-2\tilde{\Delta}_3(\aalpha)/3\log \ell+{\rm const.}$ In the plot on the right side of Fig.~\ref{fig:charged-moment-n2}, the symbols represent the coefficient $\tilde{\Delta}_3(\aalpha)$ that we get in the fit in terms of $\alpha$ for different couplings $(h=1, \gamma)$ and the curve is the CFT prediction of Eq.~\eqref{eq:scaling-dimension-result-summary} using the roots~\eqref{eq:roots_n_3}. The agreement is excellent.

\subsection{Asymptotic behavior of the entanglement asymmetry}

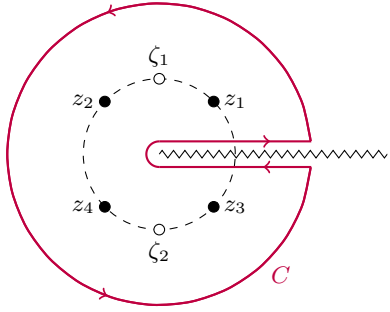
\begin{figure}[t]
\centering \small
\begin{tikzpicture}
 \draw[decoration = {zigzag,segment length = 1.5mm, amplitude = 0.5mm}, decorate] (0,0)--(3,0);
 \draw[black,dashed] (0,0) circle (1cm);
 \draw [smooth, purple,thick,domain=5:355, decoration={markings, mark=at position 0.3 with {\arrow{>}}},
        postaction={decorate}] plot ({2*cos(\x)}, {2*sin(\x)});
 \draw [thick, purple, decoration={markings, mark=at position 0.75 with {\arrow{<}}},
        postaction={decorate}] plot (0, -0.17) -- (2, -0.17);
  \draw [thick, purple, decoration={markings, mark=at position 0.75 with {\arrow{>}}},
        postaction={decorate}] plot (0, 0.17) -- (2, 0.17);
   \draw [smooth, purple,thick,domain=90:270] plot ({0.17*cos(\x)}, {0.17*sin(\x)});
\draw [smooth, purple,thick,domain=5:355, decoration={markings, mark=at position 0.7 with {\arrow{>}}},
        postaction={decorate}] plot ({2*cos(\x)}, {2*sin(\x)});
        \filldraw[black] (0.716298, 0.697795) circle (2pt);
        \filldraw[black] (-0.716298, 0.697795) circle (2pt);
        \filldraw[black] (0.716298, -0.697795) circle (2pt);
        \filldraw[black] (-0.716298, -0.697795) circle (2pt);

       \filldraw[black, fill=white] (0, 1) circle (2pt);
\filldraw[black, fill=white] (0, -1) circle (2pt);

\node at (1,0.7) {$z_1$};
\node at (-1,0.7) {$z_2$};
\node at (1,-0.7) {$z_3$};
\node at (-1,-0.7) {$z_4$};

\node at (0,1.3) {$\zeta_1$};
\node at (0,-1.3) {$\zeta_2$};

\node[purple] at (1.6,-1.6) {$C$};
\end{tikzpicture}
\caption{Schematic representation of the contour integral that gives the scaling dimension $\Delta_n(\boldsymbol{\lambda})$
for the case $n=2$. The zig-zag line is the branch cut $[0,\infty)$ of the function $\Log(z)$. The filled black dots are the 
roots $z_j$ of the polynomial $P_{\boldsymbol{\lambda}}(z)$ and the white ones represent the poles of the integrand in Eq.~\eqref{eq:Delta_contour_int}, after expanding quadratically $P_{\boldsymbol{\lambda}}(z)$ in $\llambda$.}
\label{eq:contour_int}
\end{figure}

We now compute the asymptotic behavior of the entanglement asymmetry 
for large subsystem size $\ell$ applying the general result~\eqref{eq:renyi_ent_asymm_generic}. As we have seen in Sec.~\ref{sec:setup}, for the $U(1)$ group that we are considering we have $\vol G=2\pi$, $\dim G=1$, $\mu(0)=1$, and the symmetric subgroup $H$ is the $\mathbb{Z}_2$ spin-flip symmetry. The string tension $T_n(\boldsymbol{\alpha})$ is given by Eq.~\eqref{eq:t_n_xy}, $\hess_t=t''(0)$, and, according to \cref{eq:hess_t_xy}, $t''(0)=\gamma/(2(1+\gamma))$ at the critical lines $|h|=1$. Since $\dim G =1$, the matrices $\mathsf{D}_p$, defined in Eq.~\eqref{eq:D_p}, that enter in the calculation of the coefficient $b_n$ of the $\log\ell/\ell$ term are scalars and correspond to the eigenvalues $\nu_p$ of the Hessian matrix of the scaling dimension $\Delta_n(\llambda)$, 
\begin{equation}\label{eq:hess_sc_dim_maj}
(\hess_{\Delta_n})_{ab}=\left(\frac{\partial^2{\Delta}_n(\llambda)}{\partial\lambda_a\partial\lambda_b}\right)_{\llambda=\boldsymbol{0}}, \quad a, b=1,\dots, n, 
\end{equation}
such that $\mathsf{D}_p=-2\nu_p/n$ (recall that in our case $\beta_n(\boldsymbol{\alpha})=\frac{2}{n}\Delta_n(i\boldsymbol{\alpha})$). Therefore, Eq.~\eqref{eq:renyi_ent_asymm_generic} reads in this case as

\begin{equation}\label{eq:asymmetry-asymptotic-ising}
    \Delta S_A^{(n)} = \frac{1}{2} \log \ell + a_n + b_n \frac{\log \ell}{\ell}+\cdots
\end{equation}
with
\begin{equation}
    a_n = \frac{1}{2} \log \frac{\pi t''(0) n^\frac{1}{n-1}}{2}
\end{equation}
and
\begin{equation}\label{eq:coefficient-logl/l-ising}
    b_n = \frac{1}{n(1-n) t''(0)} \sum_{p=1}^{n-1} \nu_p.
\end{equation}

\subsubsection{The Hessian of $\Delta_n({\llambda})$}

The only missing ingredient are the eigenvalues $\nu_p$ of the Hessian~\eqref{eq:hess_sc_dim_maj} of the scaling dimension $\tilde{\Delta}_n(\boldsymbol{\alpha})$. To calculate the latter, it is convenient to rewrite Eq.~\eqref{eq:scaling-dimension-result-summary} as a contour integral using the residue theorem,
\begin{equation}
    \label{eq:Delta_contour_int}
    \Delta_n(\llambda) = \frac{1}{2\pi i}\oint_C \d z f_n(z) \frac{\d}{\d z} \Log P_{\llambda} (z),
\end{equation}
with
\begin{equation}
    f_n(z) = -\frac{n^2-1}{48 \, n} + \frac{n}{4}   \left( \frac{i \Log (-z) }{2\pi}  \right)^2 . 
\end{equation}
The polynomial $P_{\boldsymbol{\lambda}}(z)$ is defined in 
Eq.~\eqref{eq:polynomial-n-defects2}. The contour $C$ encircles all the roots of $P_{\boldsymbol{\lambda}}(z)$ as we depict in Fig.~\ref{eq:contour_int}, leaving the branch cut of $\Log(z)$ outside of the region that it delimits. The advantage of this approach is that we can easily calculate the second derivatives of $\Delta_n(\boldsymbol{\lambda})$  at $\boldsymbol{\lambda}=\boldsymbol{0}$ by expanding quadratically the polynomial $P_{\boldsymbol{\lambda}}(z)$ around this point. If we rewrite~\eqref{eq:polynomial-n-defects2} in the form
\begin{equation}
    P_{\boldsymbol{\lambda}} (z) = z^n  \left\{ 2 + {\tr} \left[ R_{\lambda_1} 
    \begin{pmatrix}
        1/z & 0 \\
        0 & z
    \end{pmatrix} 
    \dots R_{\lambda_n} 
    \begin{pmatrix}
        1/z & 0 \\
        0 & z
    \end{pmatrix} 
    \right]  \right\},
\end{equation}
then it is easy perform the expansion,
\begin{equation}
    P_{\llambda}(z) = P_0(z) \slr{1 +  \frac{1}{2}  \sum_{a,b=1}^n \lambda_a \lambda_b \frac{ z^{2n-2|b-a|} + z^{2|b-a|}}{(z^n+1)^2} } + O(\lambda^3),
\end{equation}
where $P_0 (z) =  (z^n + 1)^2$. If we plug it in the contour integral~\eqref{eq:Delta_contour_int} and we integrate by parts, we find
\begin{equation}
 \Delta_n(\llambda) =  -\frac{1}{2}  \sum_{a,b=1}^n  \lambda_a \lambda_b\oint_{C} \frac{{\rm d}z}{2\pi i}  \frac{{\rm d} f_n(z)}{{\rm d}z} \frac{z^{2(n-|b-a| )} + z^{2|b-a|}}{(z^n+1)^2}+ O(\lambda^3),
\end{equation}
and
\begin{equation}
\frac{{\rm d} f_n(z)}{{\rm d}z} = - \frac{n}{8 \pi^2} \frac{\Log(-z)}{ z}.
\end{equation}
Observe that, according to this result, $\Delta_n(\boldsymbol{\lambda})=0$, as expected. Therefore, the components of the Hessian of $\Delta_n(\boldsymbol{\lambda})$
can be identified with 
\begin{equation}
    \label{eq:Hessian_intermediate}
    \lr{\mathsf H_{\Delta_n}}_{ab} =-\oint_{C} \frac{{\rm d}z}{2\pi i}  \frac{{\rm d} f_n(z)}{{\rm d}z} \frac{z^{2(n-|b-a| )} + z^{2|b-a|}}{(z^n+1)^2}.
\end{equation}
Given that $0 \leq |b-a| \leq n-1$, the numerator of the 
integrand above is, up to the ${\rm Log}(z)$ factor, a 
polynomial. Since the cut of the logarithm lies
outside the region enclosed by $C$, then the only 
singularities the contribute to the integral are the zeros of 
$z^n+1$ in the denominator, $\zeta_j = e^{i \frac{2\pi}{n} (j -\frac{1}{2}) }$, $j=1,\dots, n$. Applying the residue 
theorem, we have
\begin{equation}
    \label{eq:Hessian_intermediate_2}
     \lr{\mathsf H_{\Delta_n}}_{ab} = \frac{n}{8\pi^2} \sum_{j=1}^n \res \slr{
        \frac{\Log(-z)}{z}  \frac{z^{2(n-|b-a|)} + z^{2|b-a|} }{(z^n+1)^2} , ~\zeta_j } .
\end{equation}
These residues can be evaluated explicitly,
\begin{equation}
 \nonumber   \res \slr{ \frac{z^{p-1} \Log(-z)}{(z^n+1)^2}, ~\zeta_j }=\frac{\zeta_j^{p}}{n^2}  \left( 1 +   (p-n)  \frac{i\pi (2j-n-1)}{n}  \right) .
\end{equation}
After summing them in Eq.~\eqref{eq:Hessian_intermediate_2}, we 
eventually find that the Hessian of $\Delta_n(\boldsymbol{\alpha})$ 
is a circulant matrix,
\begin{equation}
     \lr{\mathsf H_{\Delta_n}}_{ab} = c_{a-b},
    \quad {\rm with} \quad c_l=\frac{1}{4\pi^2}\times
    \begin{cases}
     1, &\quad {\rm if}\quad l=0,\\
     \frac{2\pi}{n}\frac{l-n/2}{\sin(\frac{2\pi}{n}l)},&\quad {\rm if}\quad l=1,\dots, n-1,
    \end{cases}
\end{equation}
as a consequence of the symmetry under the cyclic exchange of the replicas.

\subsubsection{Application to the asymmetry}

According to \cref{eq:coefficient-logl/l-ising}, the coefficient of the $\log \ell/\ell$ term in the asymmetry is given by the eigenvalues $\nu_p$ of the Hessian $\mathsf H_{\Delta_n}$. In our case, since the it is a circulant matrix, the eigenvalues are given by the Fourier transform of its entries,
\begin{equation}\label{eq:nu_p}
\nu_p=\sum_{l=0}^{n-1}c_l e^{i\frac{2\pi p l}{n}}.
\end{equation}
Combining \cref{eq:coefficient-logl/l-ising,eq:nu_p}, and doing carefully the sums, we find that 
\begin{equation}\label{eq:coeff_logl_l_integers}
 b_n=\frac{\gamma+1}{2\gamma}\times\begin{cases}
        \displaystyle
        \frac{1}{(1-n)\pi^2}+\frac{1}{n(n-1)\pi}\sum_{l=1}^{n/2-1}\csc\left(\frac{2\pi l}{n}\right),&  n \text{ even}, \\
        \displaystyle
        -\frac{1}{n\pi^2}-\frac{1}{n^2(n-1)\pi}\sum_{l=1}^{n-1}(n-2l)\csc\left(\frac{2\pi l}{n}\right), &  n \text{ odd},
     \end{cases}
\end{equation}
where we have taken into account that $t''(0)=\gamma/(2(\gamma+1))$.
This result can be analytically continued to non integer values of $n$ by using the integral representation
of the cosecant function
\begin{equation}
    \csc (\pi z) = \frac{1}{\pi} \int_0^\infty \d t\frac{x^z}{x^2 + x}.
\end{equation}
Applying it in Eq.~\eqref{eq:coeff_logl_l_integers} for $n$ even, we find
\begin{equation}
    \label{eq:bn2}
 b_n=-\frac{\gamma+1}{2\gamma}\left[\frac{1}{(n-1)\pi^2}+\frac{1}{n(n-1)\pi^2} \int_0^\infty \frac{\d x}{x(x+1)} \frac{x^{2/n}-x}{1-x^{2/n}}\right].
\end{equation}
It turns out this expression reproduces the exact values of the coefficient $b_n$ for $n$ odd as well, so Eqns.~\eqref{eq:coeff_logl_l_integers} and (\ref{eq:bn2}) are equivalent expressions for all integer $n$.

Eqns.~\eqref{eq:coeff_logl_l_integers}-(\ref{eq:bn2}) are the third main result of this paper: we have arrived at the exact expression for the coefficient $b_n$ of the $\log \ell/\ell$ term in the R\'enyi entanglement asymmetry of the XY spin chain at criticality. 

Finally, taking the replica limit $n\to 1$ in Eq.~(\ref{eq:bn2}), we find that the coefficient for the (von Neumann) entanglement asymmetry is
\begin{equation}
\lim_{n\to1} b_n = -\frac{4+\pi^2}{16\pi^2} \frac{\gamma+1}{\gamma}  .
\end{equation}
Thus, our final result is that the entanglement asymmetry at criticality in the XY spin chain is
\begin{equation}
\Delta S_A =\frac{1}{2}\log\ell+\frac{1}{2}\log\frac{\pi \gamma}{4(1+\gamma)}+\frac{1}{2}-\frac{4+\pi^2}{16\pi^2} \frac{\gamma+1}{\gamma} \frac{\log\ell}{\ell}+\cdots
\end{equation}
We stress once again that what is remarkable here is the $\log \ell /\ell$ term, which only appears in critical systems. 
We also stress that the  `semi-universality' of $b_n$ (in the sense of Ref.~\cite{Stephan:2013logarithmic}) is manifest here, because it depends on the parameter $\gamma$ of the XY Hamiltonian. 
A truly universal quantity ---such as, for instance, the scaling dimension $\Delta_n({\pmb \lambda})$--- would depend only on the CFT data and not on the details of the underlying microscopic model, so it would not depend on $\gamma$.

\section{Conclusions}\label{sec:conclusions}

In this paper, we have analyzed the entanglement asymmetry in one 
dimensional critical extended quantum systems using CFT methods. This 
observable measures how much a symmetry is broken in a part of the 
system. Applying the replica trick, it can be obtained from the charged moments of the 
subsystem's reduced density matrix. We have seen that, in the ground 
state of a 1+1 dimensional quantum field theory, using the 
correspondence between the unitary operators that represent 
the symmetry group in the Hilbert space and defect lines in the path 
integral approach, the charged moments can be identified with a quotient
of the partition functions of the theory on a Riemann 
surface with and without defect lines inserted along each branch cut. 
When the state respects the symmetry, 
the defects are topological and any deformation leaves the 
partition function invariant, yielding a zero asymmetry.
In this formulation, the entanglement asymmetry can be interpreted as 
a measure of how much the defects are non topological. Utilizing well-known 
scaling arguments for the partition function in two dimensions,
we have deduced the asymptotic behavior of the charged moments that provide the 
entanglement asymmetry. While for non critical systems the moments 
decay exponentially with the subsystem size, see Refs.~\cite{rkacmb-23, Murciano:asymmetryXY, CapizziVitaleMPS:2023}, 
in the critical case we have found that they contain an extra 
algebraic factor. The coefficient of the exponential decaying term 
can be interpreted as the line tension of the defects and is 
non-universal; that is, it depends on the specific lattice realization of the field theory. The exponent 
of the algebraic factor is universal and, therefore, it is fully determined by the CFT that
describes the critical point and depends on the properties of the defects associated with the symmetry group. From this result, we have 
derived the asymptotic behavior of the ground state entanglement asymmetry for a generic compact Lie group.
Both for non-critical and critical systems, it grows at leading order logarithmically with the subsystem
size $\ell$ and a coefficient proportional to the dimension of the Lie group. Criticality
yields a $\log\ell/\ell$ correction, which is semi-universal as its coefficient depends not only on
the universal exponent of the charged moments but also on the defect tension. 

In the rest of the paper, we have specialized to the ground state of 
the XY spin chain, which explicitly breaks the $U(1)$ symmetry of spin rotations around the transverse axis. The charged moments and the entanglement asymmetry of this 
model have been investigated outside the critical lines in Ref. \cite{Murciano:asymmetryXY} 
employing lattice methods. Here we have considered the critical lines 
described by the massless Majorana fermion theory in the scaling 
limit, after fermionizing it with a Jordan-Wigner transformation. In 
this case, the defect lines correspond to a marginal deformation of 
this CFT. Exploiting conformal invariance, the universal exponent
that appears in the charged moments can be identified with the ground 
state energy of the massless Majorana fermion theory on a circle with 
equi-spaced marginal point defects of different strength.
To obtain it, we have carefully diagonalized its Hamiltonian for an arbitrary 
number of defects. Combining this result with those found in Ref. 
\cite{Murciano:asymmetryXY} for the non-universal exponential term, 
we have obtained an analytic expression for the entanglement 
asymmetry. 

A crucial point in our problem is that the defects we are considering are marginal, which
makes non-trivial the dependence of the CFT partition function on them. As we have already emphasized, 
the partition function hinges on the specific CFT and symmetry group under study. Therefore, it would be
desirable to consider other models and symmetries; for example, the $SU(2)$ group of spin rotations 
in the critical XXZ spin chain, whose continuum limit is the massless compact boson. The correspondence between global symmetries and (topological) defect lines that we exploit here can be enlarged to encompass higher-form symmetries~\cite{gksw-15}, symmetries generated by extended operators supported not only on lines but also on higher dimensional manifolds, and non-invertible symmetries~\cite{bt-18, yin-19}, which lack of an inverse element. It would be interesting to explore if the notion of entanglement asymmetry can be extended to these generalized symmetries.

\section*{Acknowledgements}
We are grateful to L. Capizzi, A. Foligno, D. X Horv\'ath, S. 
Murciano, F. Rottoli, H. Saleur, and J. M. St\'ephan for fruitful 
discussions. PC, FA and MF acknowledge support from the European 
Research Council (ERC) under Consolidator Grant NEMO No. 771536. JD 
acknowledges support from `Lorraine Universit\'e d'Excellence' 
program and from Agence Nationale de la Recherche through ANR-22-
CE30-0004-01 project ‘UNIOPEN’.

\appendix

\section{Fermionization and continuum limit of the XY spin chain}
\label{app:continuum-limit-majorana}
The fermionization of the XY spin chain is performed with the Jordan-Wigner transformation~\cite{lsm-61}. For completeness, we present it it in multiple steps. First we map the spin chain to a model with complex lattice fermions, the Kitaev chain. Then we further map the system to lattice Majorana fermions. Finally, we take the continuum limit and get the Majorana CFT. 

We consider the XY spin chain, writing explicitely the ferromagnetic coupling $J$. While in the main text this is set to $J=1$, here is important to have it in order to perform the continuum limit carefully. The Hamiltonian then is
\begin{equation}
    H_\mathrm{XY} = - \frac{J}{2} \sum_{j \in \ZZ} \lr{ \frac{1+\gamma}{2} \sigma^x_j \sigma^x_{j+1} + \frac{1-\gamma}{2} \sigma^y_j \sigma^y_{j+1} + h \sigma^z_j }.
\end{equation}

\paragraph{Kitaev chain. } 
The Jordan-Wigner transformation is 
\begin{align}\label{eq:jordan_wigner}
    \sigma^x_j &= e^{i\pi \sum_{l<j} c^\dagger_j c_j} \lr{ c^\dagger_j + c_j },  & \sigma^x_j &= e^{i\pi \sum_{l<j} c^\dagger_j c_j} i \lr{ c^\dagger_j - c_j }, & \sigma^z_j = 1 - 2 c^\dagger_j c_j,
\end{align}
and the operators $ c_j, \, c^\dagger_j$ satisfy the anticommutation relations $\{ c_j, c_l \} = 0$ and  $\{ c_j, c^\dagger_l \} = \delta_{jl}$. Then the Hamiltonian \eqref{eq:XY_ham} of the XY model is mapped to the Hamiltonian of the Kitaev chain
\begin{equation}\label{eq:Hamiltonian-Kitaev}
    H_\mathrm{Kit} = - \frac{J}{2} \sum_{j\in\ZZ} \slr{ c^\dagger_j c_{j+1} + c^\dagger_{j+1} c_j + \gamma \lr{ c^\dagger_j c^\dagger_{j+1} + c_{j+1} c_j } + h \lr{ 1-2 c^\dagger_j c_j } }
\end{equation}
and the charge $Q_A$ for $A = \{ 1, \dots, \ell \}$ becomes $Q_A =  \sum_{j \in A} c^\dagger_j c_j$.

It is well known that the XY model flows in the infra-red to the Ising CFT for $h=1, \, \gamma \in \RR$. Thus all these points in the parameter space belong to the same universality class. For simplicity, we consider the case $h=\gamma=1$. 

\paragraph{Majorana Chain.}   Each pair of complex fermion operators $c_j, c^\dagger_j$ can be split in the following pair of  Majorana operators
\begin{align}
    c^\dagger_j &= \frac{1}{2} \lr{ a_{2j} + i a_{2j+1} },
    &
    c_j &= \frac{1}{2} \lr{ a_{2j} - i a_{2j+1} },
\end{align}
which satisfy the algebra $a_j^\dagger = a_j$  and  $\{ a_j, a_j \} = 2\delta_{jl}$. Then the Hamiltonian  becomes
\begin{equation}
    H_\mathrm{Maj} = \frac{iJ}{2} \sum_{j\in\ZZ} a_{j+1} a_{j} 
\end{equation}
and the charge 
\begin{equation}
    Q_A =  \frac{1}{2} \sum_{j\in A} \lr{ 1 + i a_{2j+1} a_{2j} } .
\end{equation}

\paragraph{Continuum limit.} The continuum limit is performed by first defining the following new Majorana lattice operators $\psi_j, \, \bar\psi_j$
\begin{align}
    a_{2j} &=  \psi_j + \bar\psi_j,   & a_{2j+1} &=  \psi_j - \bar\psi_j, 
\end{align}
which form two anticommuting families of Majorana fermions, with algebra
\begin{align}
    \{ \psi_j, \psi_l \} &=  \delta_{jl}, & \{ \bar\psi_j, \bar\psi_l \} &= \delta_{jl}, & \{ \psi_j, \bar\psi_l \} &= 0.
\end{align}
The Hamiltonian in these variables reads
\begin{equation}
    H_\mathrm{Maj'} = \frac{iJ}{2} \sum_{j\in\ZZ} \lr{ 2 \psi_j \bar\psi_j + \psi_j\psi_{j-1} + \bar\psi_j \psi_{j-1} - \psi_j \bar\psi_{j-1} - \bar\psi_j \bar\psi_{j-1} }
\end{equation}
and the charge
\begin{equation}\label{eq:charge-majorana-before-continuum}
    Q_A = \frac{1}{2} \sum_{j\in A} \lr{ 1 + i 2 \psi_j \bar\psi_j }.
\end{equation}
Finally, we perform the continuum limit. We call the continuum variable $x \in \RR$ and introduce a lattice spacing $s$ so that 
\begin{align}
    \psi_j &\simeq \frac{\psi(x)}{\sqrt{s}}, & \psi_{j-1} &\simeq \frac{\psi(x-s)}{\sqrt{s}} \simeq \frac{1}{\sqrt s} (\psi(x) - s \partial_x \psi(x)).
\end{align}
The continuum fields satisfy the algebra $\{ \psi(x), \psi(y) \} = \delta(x-y), \, \{ \bar\psi(x), \bar\psi(y) \} = \delta(x-y), \, \{ \psi(x), \bar\psi(y) \} = 0$. The Hamiltonian becomes
\begin{equation}
    H = \frac{J'}{2i} \int_\RR \d x \slr{ \psi(x) \partial_x \psi(x) - \bar\psi(x) \partial_x \bar\psi(x) },
\end{equation}
where $J'$ is the continuum version of $J$, given by $J' = J s$ in the limit $s \to 0$ and $J \to \infty$. Deriving the equations of motion for $\psi$ and $\bar\psi$, $J'$ can be recognized to be the sound velocity, which we set to $1$. Finally, the charge operator, discarding the constant term in \cref{eq:charge-majorana-before-continuum} that acts trivially on the Hilbert space, becomes
\begin{equation}
    Q_A = i \int_A \psi(x) \bar\psi(x) \d x.
\end{equation}

\section{Defects in the Hamiltonian formalism}\label{app:defect_Ham_formalism}

In this Appendix, we consider a massless Majorana fermion on a line, with a defect implemented as a localized mass term
\begin{equation}
	 H_\mu \, = \, \frac{1}{2 i} \int_\RR  (   \psi \partial_x \psi   -  \bar{\psi} \partial_x \bar{\psi} ) \d x \, -\,  i \mu \,\psi (0 )\bar{\psi}(0), \qquad \mu \in \RR.
\end{equation}
We show that this formulation is equivalent to the one given in the main text, where the defect is only encoded in the gluing conditions. We provide the explicit relation between the defect strength $\mu$ and the gluing parameter $\lambda$. We find that if the defect term in the Hamiltonian is Hermitian, then $\lambda$ has to be real. This is a further justification of the analytic continuation $\alpha \mapsto -i \lambda$ that is performed in the main text.

We have the following commutators
\begin{align}
	\left[ H_\mu, \psi(x) \right]  & =   i \partial_x \psi(x)  +  i   \mu  \delta(x)  \bar{\psi}(0), \\
	\left[ H_\mu , \bar{\psi}(x) \right]  & =  - i   \partial_x \bar{\psi}(x)  -   i  \mu \delta(x)  \psi(0) .
\end{align}
To relate this to gluing conditions at the origin, we can look for eigenmodes of that Hamiltonian of the form
\begin{equation} \label{eq:etak}
    \eta_k = \int_\RR \slr{ u^*_k (x) \psi(x) + v^*_k(x) \bar \psi(x) } \d x,
\end{equation}
with 
\begin{align}
    u_k(x) &=
    \begin{cases}
        A_0 e^{ikx}, & x < 0, \\
        A_1 e^{ikx}, & x > 0, 
    \end{cases}
    &
    v_k(x) &=
    \begin{cases}
        B_0 e^{-ikx}, & x < 0, \\
        B_1 e^{-ikx}, & x > 0, 
    \end{cases}
\end{align}
for some constants $A_0, \, A_1, \, B_0, \, B_1$. This Ansatz gives the following commutator with the Hamiltonian,
\begin{equation}
    [H_\mu, \eta_k] = k \eta_k - i \lr{ A_0^* - A_1^* - \frac{\mu}{2} (B_0^* + B_1^*) } \psi(0)  + i  \lr{ B^*_0 -  B^*_1  - \frac{\mu}{2} (A^*_0+A^*_1)} \bar{\psi}(0). 
\end{equation}
We see that $\eta_k$ is a Bogoliubov mode with energy $k$ if the last two terms vanish. This gives the constraint
\begin{equation}
	\begin{pmatrix}
	   A_0 \\ B_0 
    \end{pmatrix}
    =
    \begin{pmatrix}
		\frac{1+ \mu^2/4}{1-  \mu^2/4}  &   \frac{\mu}{1- \mu^2/4} \\
		\frac{\mu}{1- \mu^2/4}  & \frac{1+ \mu^2/4}{1-  \mu^2/4}
	\end{pmatrix}
    \begin{pmatrix}
        A_1 \\ B_1
    \end{pmatrix} .
\end{equation}
Thus, we recover the gluing condition (\ref{eq:gluing-cylinder}) with the matrix (\ref{eq:gluing-matrix-lambda}) obtained after the analytic continuation of the gluing parameter, provided that
\begin{equation}
    \frac{\mu}{2} = \tanh \left(\frac{\lambda}{2} \right) .
\end{equation}

\section{Numerical calculation of the charged moments}
\label{app:numerics}

In this Appendix, we report the formulae that we employ to compute 
numerically the charged moments~\eqref{eq:def-Z(g)} for the $U(1)$ 
group of spin rotations around the $z$ axis in the ground state of 
the XY spin chain~\eqref{eq:XY_ham}. As we show in Appendix~\ref{app:continuum-limit-majorana}, this model maps into a quadratic fermionic chain after the Jordan-Wigner
transformation~\eqref{eq:jordan_wigner}. Therefore, its ground state satisfies Wick theorem. This implies that the reduced
density matrix $\rho_A$ of a single interval $A$  of length $\ell$ is Gaussian and it is fully determined by the 
$2\ell\times 2\ell$ two-point fermionic correlation matrix~\cite{peschel-02}
\begin{equation} 
 \Gamma_{jj'}=2\mathrm{tr}\left[\rho_A
 \left(\begin{array}{c} c_j \\ c_j^\dagger\end{array}\right)
 (c_{j'}^\dagger, c_{j'})\right]-\delta_{jj'},
\end{equation}
with $j,j'=1,\dots,\ell$. For the ground state of the XY spin chain, its entries are
\begin{equation}\label{eq:Gammat0}
\Gamma_{jj'}=\int_{0}^{2\pi}\frac{{\rm d}k}{2\pi}\mathcal{G}(k)e^{-ik(j-j')},
\end{equation}
where $\mathcal{G}(k)$ is the $2\times 2$ matrix 
\begin{equation}\label{eq:gs_symbol}
    \mathcal{G}(k)=\left(\begin{array}{cc}
    \cos\xi_k & -i \sin\xi_k\\
     i\sin\xi_k& -\cos\xi_k
    \end{array}\right)
\end{equation}
and $\cos\xi_k$, $\sin\xi_k$ are given in Eq.~\eqref{eq:bogo_angle}.

After the Jordan-Wigner transformation, the transverse magnetization~\eqref{eq:transv_mag} that generates the $U(1)$ symmetry
is also quadratic and, consequently, Gaussian. Therefore, the 
charged moments $\mathcal{Z}_n(\aalpha)$ are the trace of a 
product of Gaussian operators. Using the well-known properties of 
this kind of operators, the charged moments can be calculated 
in terms of the two-point correlataion matrix $\Gamma$ as
\begin{equation}\label{eq:numerics}
  \mathcal{Z}_n(\boldsymbol{\alpha})=\sqrt{\det\left[\left(\frac{I-\Gamma}{2}\right)^n
  \left(I+\prod_{j=1}^n W_j\right)\right]},
\end{equation}
where $W_j=(I+\Gamma)(I-\Gamma)^{-1}e^{i\alpha_{j,j+1} n_A}$ and $n_A$ is a diagonal matrix with $(n_A)_{2j,2j}=1$, $(n_A)_{2j-1,2j-1}=-1$, $j=1, \cdots, \ell$. The detailed derivation of this expression can be found in Ref.~\cite{amvc-23}.
We use it to obtain the exact numerical values of the charged moments in the plots of Fig.~\ref{fig:charged-moment-n2}.

\end{document}